\documentclass[twocolumn]{aastex631}

\newcommand{\htwo}{\ensuremath{{\rm H}_2}}
\newcommand{\hsix}{\ensuremath{{\rm H}_6^+}}
\newcommand{\hdthree}{\ensuremath{{\rm (HD)}_3^+}}
\newcommand{\hminus}{\ensuremath{{\rm H}^-}}

\newcommand{\ket}[1]{\ensuremath{| #1\rangle}}
\newcommand{\element}[3]{\ensuremath{\langle #1|\,#2\,| #3\rangle}}
\newcommand{\axx}{\ensuremath{\alpha_{_{\!X\!X}}}}
\newcommand{\azz}{\ensuremath{\alpha_{_{\!Z\!Z}}}}
\newcommand{\exxxx}{\ensuremath{{\rm E}_{_{\!X\!,X\!X\!X}}}}
\newcommand{\ezzzz}{\ensuremath{{\rm E}_{_{\!Z\!,Z\!Z\!Z}}}}
\newcommand{\bxxxx}{\ensuremath{{\rm B}_{_{\!X\!,X\!,X\!X}}}}
\newcommand{\bxxzz}{\ensuremath{{\rm B}_{_{\!X\!,X\!,Z\!Z}}}}
\newcommand{\bxzxz}{\ensuremath{{\rm B}_{_{\!X\!,Z\!,X\!Z}}}}
\newcommand{\bzzzz}{\ensuremath{{\rm B}_{_{\!Z\!,Z\!,Z\!Z}}}}
\newcommand{\cxxxx}{\ensuremath{{\rm C}_{_{\!X\!X\!,X\!X}}}}
\newcommand{\cxzxz}{\ensuremath{{\rm C}_{_{\!X\!Z\!,X\!Z}}}}
\newcommand{\czzzz}{\ensuremath{{\rm C}_{_{\!Z\!Z\!,Z\!Z}}}}
\newcommand{\gxxxx}{\ensuremath{\gamma_{_{\!X\!X\!X\!X}}}}
\newcommand{\gxxzz}{\ensuremath{\gamma_{_{\!X\!X\!Z\!Z}}}}
\newcommand{\gzzzz}{\ensuremath{\gamma_{_{\!Z\!Z\!Z\!Z}}}}
\newcommand\be{\begin{equation}}
\newcommand\ee{\end{equation}}

\graphicspath{{./}{figures/}}

\shorttitle{Absorption spectra of electrified \htwo}
\shortauthors{Walker}

\begin{document}

\title{Absorption spectra of electrified hydrogen molecules}

\correspondingauthor{Mark Walker}
\email{mark.walker@manlyastrophysics.org}

\author[0000-0002-5603-3982]{Mark A. Walker}
\affil{Manly Astrophysics,
15/41-42 East Esplanade,
Manly, NSW 2095, Australia}

\begin{abstract}
Molecular hydrogen normally has only weak, quadrupole transitions between its rovibrational states, but in a static electric field it acquires a dipole moment and a set of allowed transitions. Here we use published \emph{ab initio} calculations of the static electrical response tensors of the \htwo\ molecule to construct the perturbed rovibrational eigensystem and its ground state absorptions. We restrict attention to two simple field configurations that are relevant to condensed hydrogen molecules in the interstellar medium: a uniform electric field, and the field of a point-like charge. The energy eigenstates are mixtures of vibrational and angular momentum eigenstates so there are many transitions that satisfy the dipole selection rules. We find that mixing is strongest amongst the states with high vibrational excitation, leading to hundreds of absorption lines across the optical and near infrared. These spectra are very different to that of the field-free molecule, so if they appeared in astronomical data they would be difficult to assign. Furthermore in a condensed environment the excited states likely have short lifetimes to internal conversion, giving the absorption lines a diffuse appearance. We therefore suggest electrified \htwo\ as a possible carrier of the Diffuse Interstellar Bands (DIBs). We further argue that in principle it may be possible to account for all of the DIBs with this one carrier. However, despite electrification the transitions are not very strong and a large column of condensed \htwo\ would be required, making it difficult to reconcile this possibility with our current understanding of the ISM.
\end{abstract}


\section{Introduction}
It is widely appreciated that cold, dense gas is difficult to see, and that this is in part because its main constituents -- i.e. helium and molecular hydrogen -- have no electric dipole transitions below the far-UV. It is therefore possible that galaxies could contain substantial, undetected reservoirs of cold, dense gas, and \citet{pfenniger1994a} argued that must be the case in order to account for the observed properties of star-forming galaxies. \citet{pfenniger1994b} noted that in such reservoirs the hydrogen ought to be close to its saturation vapour pressure and could manifest particles of solid \htwo\ --- an idea that ultimately led to a revival of interest in solid \htwo\ as a candidate material for interstellar dust, for the reasons outlined below. 

An important aspect of condensed \htwo\ is that its ionisation chemistry differs from that of the gas phase, with the dominant ionisation product being ${\rm H}_3^+$ for gaseous \htwo\ \citep[see, e.g., the review by][]{miller2020}, and the balance shifting towards \hsix\ in condensed environments \citep[e.g.][]{jaksch2008}. As a result the \hsix\ molecule -- which was only recently revealed in the laboratory, using electron spin resonance \citep{kumada2005} -- could be abundant in the interstellar medium (ISM). The rovibrational spectra of that molecule offer a means of recognising solid hydrogen in the ISM, but have not yet been studied in the laboratory. Using \emph{ab initio} calculation \citet{lin2011} demonstrated that the two most important isotopomers, namely \hsix\ and \hdthree, both have vibrational transitions that are coincident with strong mid-infrared lines observed from the ISM, and suggested that they could be the carriers of the observed lines. 

Considering charged species such as \hsix\ inside a matrix of solid \htwo\ naturally suggested another interesting possibility \citep{lin2011}: astrophysical hydrogen ices might be substantially more robust than their laboratory counterparts, as a result of electrostatic interactions associated with ionic impurities. And interstellar ice particles, even if they were initially pure \htwo, would soon acquire charged impurities as a result of bombardment by charged particles and ionizing photons. That is a critical point because it was shown by \citet{field1969} and \citet{greenberg1969} that pure \htwo\ ice sublimates rapidly under typical interstellar conditions, and subsequent to those papers solid hydrogen was eschewed as a candidate material for interstellar dust. Consequently there is a well-developed body of theoretical work on silicate and carbonaceous interstellar dust grains \citep[e.g.,][]{draine2003}, but we know very little about the possible manifestations of \htwo\ ice particles.

The influence of collisional charging on the lifetime of hydrogen dust particles was considered in detail by \citet{walker2013} who emphasized the important role played by the unusual electronic properties of solid hydrogen. Specifically, the conduction band lies above the energy of an electron in vacuo, which makes it difficult for electrons to penetrate the lattice; instead they tend to become trapped above the surface, where they form a two-dimensional electron gas \citep{cole1970}. But there is no such barrier for ionic species, which easily penetrate into the matrix, so that grain charging can create large column-densities in charges of both signs and a grain that is close to overall neutrality. Consequently the electric fields inside the matrix may reach high values, with a correspondingly large attenuation of the \htwo\ sublimation rate and a greatly increased lifetime of solid \htwo\ particles \citep{walker2013}. In this picture the electric field is not localized and may be considered to be approximately uniform within the solid --- as an idealized example one can imagine the case of a plate-like crystal habit  \citep[such as is known to occur amongst the many types of terrestrial H$_2$O snowflakes,][]{libbrecht2017} with positive charges near the midplane and electrons bound to the surfaces.

At the opposite extreme it has long been recognized that localized electric fields can bind \htwo\ ligands, and that this can occur under conditions where pure \htwo\ would not condense. For example: \citet{sandford1993} undertook laboratory studies of the condensation of \htwo\ molecules onto water ice, which is promoted by the electric dipole moment of H$_2$O; and, \citet{duley1996} analysed the formation mechanisms of \htwo\ clusters around ionic species, in the context of dense interstellar clouds. If cation and anion clusters are both present then they can be expected to aggregate, leading to dust particles of hydrogen ice. Provided that the identity of the main impurities is known, ices of this type can in principle be recognized by the absorption spectra of the impurity species, whose lines are expected to be slightly shifted in wavelength as a consequence of perturbation by the \htwo\ ligands --- the ``matrix shift''. It was suggested by \citet{bernstein2013} that the matrix-shifted electronic transitions of a large set of impurity species could be responsible for the absorption lines known as the Diffuse Interstellar Bands (DIBs).

An absorption signature of the \htwo\ matrix itself would be valuable because it would allow the main component of a hydrogen-ice particle to be observed directly, and because it would obviate the need for a detailed understanding of the impurity content of the ice. It is known that pure solid \emph{para}-\htwo\ manifests spectral features that are absent in the gas phase \citep[e.g.][]{mengel1998}, so such a signature exists, but in practice those spectral features are much too weak to be detectable in astrophysical dust \citep{kettwich2015}. Indeed by modelling the refractive index \citet{kettwich2015} demonstrated that extinction by particles of pure solid \emph{para-}\htwo\ should be scattering-dominated at frequencies below the far-UV, and even the strongest rovibrational absorptions (i.e. those that are also present in gas phase) are too weak to be seen in the extinction curve. In turn this is, of course, because the rovibrational lines of the isolated \htwo\ molecule are electric quadrupole transitions -- with some contribution from magnetic dipole \citep{roueff2019} -- and are all very weak. However, just as the \htwo\ grain sublimation rate can be strongly influenced by the presence of electric charges, so can the absorption spectrum of the hydrogen matrix. The key point is that \htwo\ becomes infrared active in the presence of a static electric field, because the field polarizes the molecule \citep{condon1932}. An electric field also changes the energy levels of \htwo, lifting degeneracies and causing shifts and splittings in its spectral lines --- a condition we refer to as ``electrified''. The radiative transitions of electrified molecules can be thought of as the zero-frequency limit of Raman spectra, with corresponding selection rules \citep{condon1932}.

In this paper we calculate the \emph{ortho-/para-}\htwo\ ground state absorption spectra that arise in two simple field configurations: a uniform field, and the field of a point-like charge. These particular choices were motivated by consideration of the types of fields that might plausibly be encountered inside charged dust grains of hydrogen ice, as discussed above. It is clear at the outset that our adopted field configurations are simplified models of the physical circumstances under consideration because, for example, we are neglecting the field due to the induced dipole moments of neighbouring \htwo\ molecules. Nevertheless the calculated spectra are complex and interesting in their own right, and they provide useful first approximations which establish a foundation for more sophisticated treatments.

There have been numerous experimental studies of the absorption spectra of irradiated solid hydrogens \citep[e.g.][]{souers1980,brooks1985,momose2001}, in which new features have been observed and attributed to the Stark-shifted absorptions of \htwo\ (or D$_2$ etc.) molecules adjacent to charged impurities. By contrast there has been relatively little exploration of their theoretical counterpart spectra, despite the availability of suitable electrical response tensors \citep{koloswolniewicz1967,pollwolniewicz1978}. That is understandable as the strong electric fields that are encountered lead to high levels of mixing amongst the basis states, and at the time the experimental studies were being pursued the task of diagonalizing the Hamiltonian matrix was computationally challenging. The principal theoretical resource to date has been the work of \citet{pollhunt1985} who quantified the energy levels of sixteen low-lying rovibrational states of \htwo, D$_2$, and T$_2$, as a function of the separation between the molecule and a point-like charge of either sign. In \S3.3 we give a quantitative comparison between their results, for the case of \htwo, and ours.

There are some qualitative differences between the treatment of \citet{pollhunt1985} and the work presented here. First, because we are concerned with the astrophysical context we have not attempted to characterize the states of any of the heavier isotopes of hydrogen. Secondly, our calculations are not restricted to a few of the low-lying eigenstates: all of the perturbed eigenstates are characterized, because it is now computationally feasible to do so. Thirdly, we evaluate the electric dipole transition matrix elements between all pairs of rovibrational states, allowing us to determine the transition strength and natural width appropriate to every line in the  absorption spectrum of the electrified molecule. Finally, we extend the electrical response model of \htwo\ from the two second-rank tensors -- the permanent quadrupole, $\Theta$, and the dipole polarizability, $\alpha$ -- that were considered by \citet{pollhunt1985} to include five response tensors of fourth rank. Our expansion in scope of the electrical response model was made possible by the recent \emph{ab initio} calculations of \citet{miliordos2018}, whose treatment of \htwo\ is comprehensive up to fourth rank.

The structure of this paper is as follows. In the next section we set out the inputs needed for our calculations, namely the eigenstates of the unperturbed \htwo\ molecule and the perturbation introduced by a static electric field. Some details of the latter are placed in appendices as they are required for calculations but not essential for understanding. Section 3 describes the eigenstates of the perturbed molecule, which we obtain by diagonalising the perturbed Hamiltonian, for three distinct cases: a uniform electric field; a point-like elementary positive charge; and, a point-like elementary negative charge. In each case the eigensystem is a function of a single parameter (e.g. field strength). Section 4 presents our calculated absorption spectra of ground-state molecules for each of these cases; both \emph{ortho-} and \emph{para-} sequences are treated, so there are two ground states and two absorption spectra for every field configuration. Section 5 considers how our calculated spectra might relate to the observed properties of the ISM, with particular emphasis on the possibility that electrified \htwo\ might be a carrier of the DIBs. Then in the Discussion section (\S6) we detail various issues arising with that idea, one of which is of particular importance because it is a potential show-stopper: the strength of the transitions of electrified \htwo. The problem, discussed in \S6.4, is that trace quantities of electrified \htwo\ cannot explain the DIBs --- a substantial fraction of the total interstellar mass would need to be in the form of condensed hydrogen, implying substantial revisions to our understanding of the ISM. Section 6 also provides some commentary on the limitations of the theoretical framework that we have employed. Summary and conclusions follow in \S7.

\section{Description of the unperturbed molecule}
In the present paper we will be concerned exclusively with the electronic ground state, X$\,^1\Sigma_g^+$, of the \htwo\ molecule; although that restriction will not be restated from here onwards it should be understood to apply throughout.

The \htwo\ molecule has been well studied both experimentally and theoretically. It is known that there are 302 bound rovibrational levels of \htwo\ \citep{komasa2011}, characterized by two quantum numbers: vibrational, $v$, with $0\le v\le 14$, and total angular momentum, $j$, with $0\le j\le 31$. Each level is a multiplet of $2j+1$ degenerate states distinguished from each other by one component of the angular momentum -- which we take to be the $z$-component -- and labelled with the corresponding quantum number, $m$, where $-j\le m \le j$. The total number of bound rovibrational states \ket{v,j,m} is thus $7{,}086$; this set of states constitutes the basis that spans our modelling space. Unfortunately there is no guarantee that this basis suffices to describe the bound states of the perturbed molecule -- in general the unbound continuum states may also be required -- and in \S4.2.3 we draw attention to evidence for incompleteness in the case of strong electric fields. 

Theoretical level energies are known to high precision and compare favourably with the available laboratory spectroscopy, with estimated accuracy better than $0.001\,{\rm cm^{-1}}$ \citep{komasa2019}. Although the level energy calculations themselves are naturally done in atomic units, it is conventional in laboratory spectroscopy to use the wavenumber, $1/\lambda$, where $\lambda$ is the vacuum wavelength of the transition in ${\rm cm}$; we follow that convention throughout this paper. 

The level energies are, of course, just the eigenvalues of the rovibrational Hamiltonian matrix, so adopting the \ket{v,j,m} states as our basis, and using the energy eigenvalues reported by \citet{komasa2011}, we can immediately construct the unperturbed Hamiltonian as a diagonal matrix whose entries are those eigenvalues. In principle that construction could include all the rovibrational states of \htwo\ but in practice it is better to split the calculation in two, considering \emph{ortho}-\htwo\ ($j=1,3,5...$) and \emph{para}-\htwo\ ($j=0,2,4...$) separately. As is well known, these two sequences can usually be considered as distinct species because the \emph{ortho} states all have parallel nuclear spins (nuclear spin triplet), while the \emph{para} states all have antiparallel spins (nuclear spin singlet), and the rate of interconversion is very small \citep{freiman2017}. We will see that the perturbations introduced by a static electric field do not mix \emph{ortho} states with \emph{para} states, so it is valid to consider the two sequences separately. Doing so has computational advantages because it is quicker to construct and diagonalize the two separate Hamiltonians than it is to handle everything in a single matrix.

\subsection{Unperturbed states}
To determine the energy eigenstates of the perturbed molecule we need to calculate the Hamiltonian matrix elements of the perturbation, and in order to evaluate those elements we must first characterize the unperturbed states themselves. In other words we must solve the time-independent Schr\"odinger equation for the unperturbed molecule. To do so we employ the Born-Oppenheimer approximation and consider only the wavefunction describing the internuclear separation $\mathbf{r}$. As usual we can separate the angular and radial variables, leading to eigenfunctions of the form $\psi(r,\theta,\phi) = Y_{j,m}(\theta,\phi)\,\chi_{v,j}(r)$, in terms of the spherical harmonics $Y_{j,m}$. The remaining differential equation describes the dependence of the wavefunctions on the radial coordinate:
\be
{{{\rm d\ }}\over{{\rm d}r}}\!\left(\! r^2 {{{\rm d}\chi_{v,j}}\over{{\rm d}r}}\! \right)+{{m_p}\over{m_e}}\,r^2 \left[E_{v,j}-V\right]\chi_{v,j}
-j(j+1)\chi_{v,j}=0,\label{eq:schrodingerequation}
\ee
with all quantities in atomic units. To solve for the various $\chi_{v,j}$ and the corresponding energy eigenvalues $E_{v,j}$ it is necessary to know the internuclear potential $V(r)$; we use the Born-Oppenheimer potential tabulated by \citet{pachucki2010} plus the adiabatic correction function given by \citet{pachuckikomasa2014}.

To obtain solutions to equation (\ref{eq:schrodingerequation}) we use the Numerov-Cooley method \citep{cooley1961,johnson1977}, which is sixth-order accurate in the step size, to solve for the function ${\cal R}_{v,j}\equiv r\,\chi_{v,j}$. Our wavefunctions were determined on a uniform grid of 20{,}000 points extending to $r=20$ (atomic units).\footnote{The atomic unit of distance is also known as the Bohr radius; the approximate numerical value is $0.529\;{\rm \AA}.$} As part of the solution process we obtain our own set of eigenvalues. Our eigenvalues differ by $2\pm 2\,{\rm cm^{-1}}$ from the much more accurate eigenvalues of \citet{komasa2011}, implying systematic errors at the level $\sim 10^{-4}$ (fractional error in the eigenvalue) for most of the eigenstates. This level of accuracy suffices for constructing the matrix elements of the electrical perturbation because it is substantially smaller than the systematic errors in the input values of the electrical response tensors (see next section). We note, however, that the level $(v,j)=(14,4)$ has such a small binding energy \citep[$0.0265\,{\rm cm^{-1}}$,][]{komasa2011} that it cannot be adequately described by our approach and we have therefore excluded it from our calculations. That leaves us with 155 energy levels and 3{,}567 possible states for \emph{para}-\htwo, and 146 energy levels and 3{,}510 states for \emph{ortho}-\htwo.

\subsection{Static electrical response of \htwo}
Recent \emph{ab initio} calculations by \citet{miliordos2018} have substantially improved the characterisation of the electrical response of the \htwo\ molecule. In addition to the second-rank response tensors -- the permanent quadrupole moment, $\Theta$, and the dipole polarizability, $\alpha$ -- \citet{miliordos2018} evaluated the permanent hexadecapole moment, $\Phi$, the quadrupole polarizability, ${\rm C}$, the dipole-octupole polarizability, ${\rm E}$, the dipole-dipole-quadrupole hyperpolarizability, ${\rm B}$, and the second dipole hyperpolarizability, $\gamma$. In other words: all of the (non-zero) response tensors up to fourth rank. And each independent component of those tensors was tabulated by \citet{miliordos2018} over a wide enough range of internuclear separations ($0.567 \le r \le 10$, atomic units) that accurate matrix elements can be evaluated between all pairs of bound rovibrational states.

Good numerical accuracy was demonstrated by \citet{miliordos2018}, with test calculations for the case of a ground state hydrogen atom yielding electrical response tensors that differ from the known (exact) values by only $\sim 0.4\%$. Although that is a small error it is large compared to the systematic errors in the eigenfunctions we have constructed (\S2.1), so we do not expect the latter to limit the accuracy of the matrix elements we have evaluated.

Because of the symmetries of the \htwo\ molecule (point group $D_{\infty h}$) only 16 independent components are required to fully describe the seven response tensors under consideration \citep{mclean1967,buckingham1967,miliordos2018}. For the convenience of readers, the behaviour of those components as a function of internuclear separation is presented in Appendix A; they are specified in a cartesian coordinate system $(X,Y,Z)$ in which the $Z$-axis is aligned with the internuclear separation vector. The response tensors in the laboratory coordinate system, $(x,y,z)$, depend also on the orientation of the internuclear axis and can be related to those in the molecular basis using the expressions given by \citet{buckingham1967}, for $\Theta$, $\alpha$, ${\rm C}$, ${\rm B}$, and $\gamma$, and by \citet{bohrhunt1987} for $\Phi$ and ${\rm E}$. We are thus in a position to evaluate the perturbation of the energy of the molecule \citep{miliordos2018}
\begin{eqnarray}
\Delta E = & -{1\over3}\Theta_{\alpha\beta}F^\prime_{\alpha\beta} 
\,-\,{1\over{105}}\Phi_{\alpha\beta\gamma\delta}F^{\prime\prime\prime}_{\alpha\beta\gamma\delta} -{1\over 2}\alpha_{\alpha\beta}F_\alpha F_\beta \hskip1.3cm \label{eq:electricalenergy}\nonumber \\ 
& -\,{1\over 6}{\rm C}_{\alpha\beta,\gamma\delta}F^\prime_{\alpha\beta} F^\prime_{\gamma\delta}
-{1\over {15}}{\rm E}_{\alpha,\beta\gamma\delta}F_{\alpha} F^{\prime\prime}_{\beta\gamma\delta}\hskip2.3cm\\ \nonumber
& -{1\over 6}{\rm B}_{\alpha,\beta,\gamma\delta}F_{\alpha}F_{\beta} F^\prime_{\gamma\delta}
\,-\,{1\over {24}}\gamma_{\alpha\beta\gamma\delta}F_{\alpha} F_{\beta} F_{\gamma} F_{\delta},\hskip 1.4cm
\end{eqnarray}
arising from the imposed electric field, $F$, and its first three gradients $F^\prime\equiv\nabla F$, etc.

\section{Electrified states of molecular hydrogen}
To determine the eigenstates of \htwo\ in the presence of an electric field we first evaluate the Hamiltonian matrix elements of the electrical perturbation in the unperturbed basis. Those elements are then added to the unperturbed Hamiltonian, which is diagonal and very accurately known, and the full Hamiltonian is diagonalized to yield the new eigenvalues and eigenvectors.

The form of the perturbation given in equation (\ref{eq:electricalenergy}) is not well suited to the task of rapid calculation of the millions of elements that make up the Hamiltonian matrix. For that purpose it is much more convenient to expand the perturbation in spherical harmonics. For the electric field configurations under consideration in this paper, both of which are axisymmetric, the only non-zero coefficients are associated with $Y_{l,0}$, with degrees $l=0,2,4$. The expansion coefficients themselves are given in Appendix B for both the uniform field case, and for the field of a point-like charge. Using those expansions, matrix elements of the perturbation are readily evaluated as the product of radial overlap integrals, which depend only on the $(v,j)$ combinations for the two states under consideration, and the angular overlap integrals which depend only on the $(j,m)$ combinations and are easily determined via the well known Wigner-3J coefficients (Appendix B).

Because each of the radial overlap integrals is used many times over -- for states with different values of $m$, for the various coefficients in the spherical harmonic expansion of the perturbation, for investigating different field configurations, and for the dipole moment operator (\S4) as well as the Hamiltonian -- it is computationally efficient to evaluate all integrals of the form
\be
\langle {\cal T}(v,j;u,k)\rangle=\int\,{\rm d}r\,{\cal R}_{v,j}(r)\,{\cal T}(r)\,{\cal R}_{u,k}(r),\label{eq:tensorcomponentexpectationvalues}
\ee
(where ${\cal T}$ stands for any of the 16 independent tensor components) and store the resulting matrices. With only $\sim 150$ different combinations of $(v,j)$ for each nuclear spin state, the required storage space is modest.

For the field configurations under study in this paper, the spherical harmonic expansion of the perturbation contains only a small number of terms so the Hamiltonian matrix is sparse and can be stored in a compact form. Much more space would be required to store the eigenvectors of the Hamiltonian, which form a dense matrix, so it is preferable to store the Hamiltonian itself; when the eigenvalues and eigenvectors are needed in future, the Hamiltonian can be read from the filesystem and diagonalized in only a few seconds.

An immediate consequence of axisymmetry in the electric field is that the perturbation only mixes together basis states having the same value of $m$, because the matrix elements  $\element{j,m}{Y_{l,0}}{j^\prime,m^\prime}$ are zero unless $m^\prime=m$. Consequently the $z$ component of the angular momentum is well-defined, and $m$ is a good quantum number of the perturbed molecule.

Because the electrical perturbation is not isotropic there is mixing of basis states with different values of the total angular momentum, and therefore $j$ is not a good quantum number for the electrified states. The fact that the angular structure of the electrical perturbation includes only spherical harmonics with even values of the degree, $l$, means that \emph{para}-\htwo\ basis states are only mixed together with other \emph{para}-\htwo\ basis states, and similarly for \emph{ortho}-\htwo. Consequently the \emph{ortho}- and \emph{para}- sequences can be analyzed separately, as noted in \S2.

Finally there is mixing of different vibrational eigenstates, because none of the off-diagonal elements of the response matrices $\langle{\cal T}\rangle$ of equation (\ref{eq:tensorcomponentexpectationvalues}) are expected to vanish, and therefore $v$ is not a good quantum number of the perturbed molecule. However, for the low-lying rovibrational states, with small $v$ and small $j$, the energy difference between adjacent vibrational states is large compared to the energy difference between adjacent rotational states and there is much less mixing of vibrational eigenstates than rotational eigenstates \citep{pollhunt1985}. 

Our calculations were performed using the \emph{Mathematica}\footnote{http://www.wolfram.com} software package, which provides a facility for obtaining the eigensystem of the Hamiltonian. Despite the large dimensions of that matrix, diagonalization requires only a few seconds on a laptop computer. 

\begin{figure}
\centering
\gridline{\fig{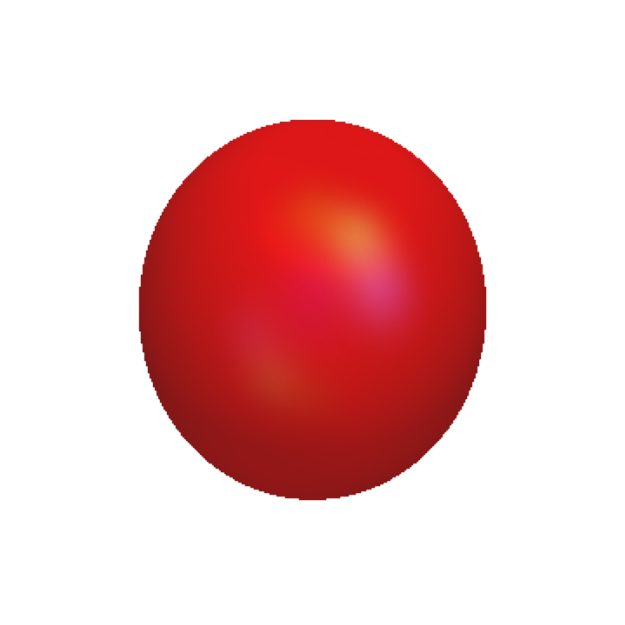}{0.15\textwidth}{}
          \fig{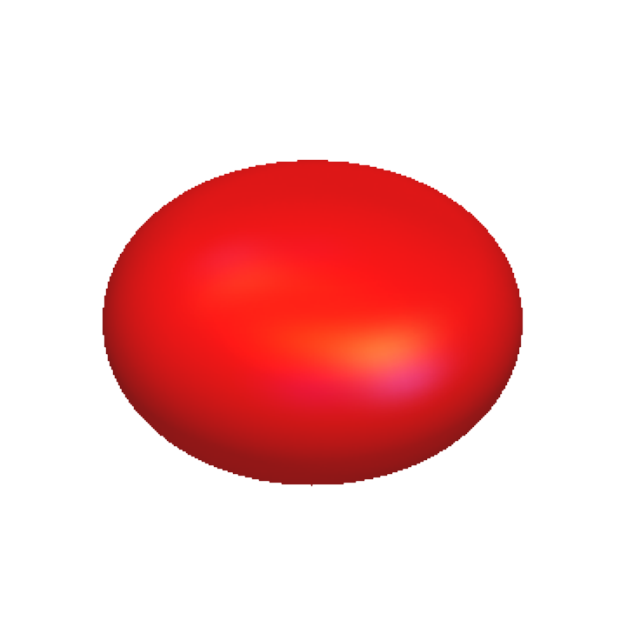}{0.15\textwidth}{}
          \fig{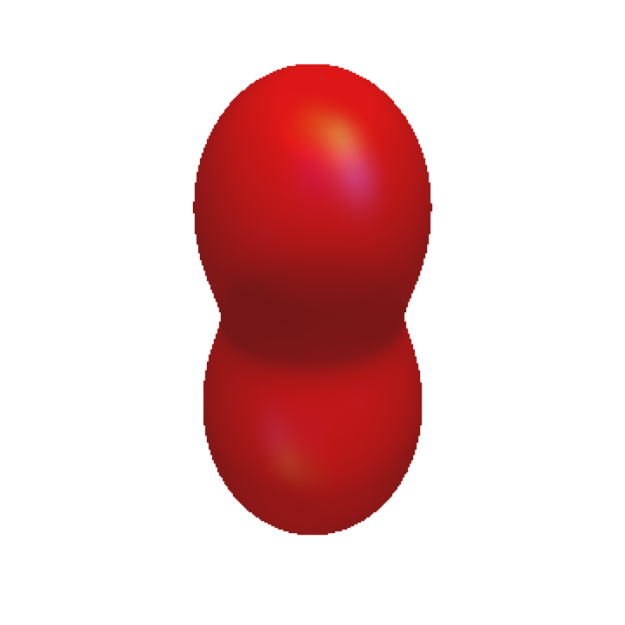}{0.15\textwidth}{}}
\vskip -1.2cm          
\gridline{\fig{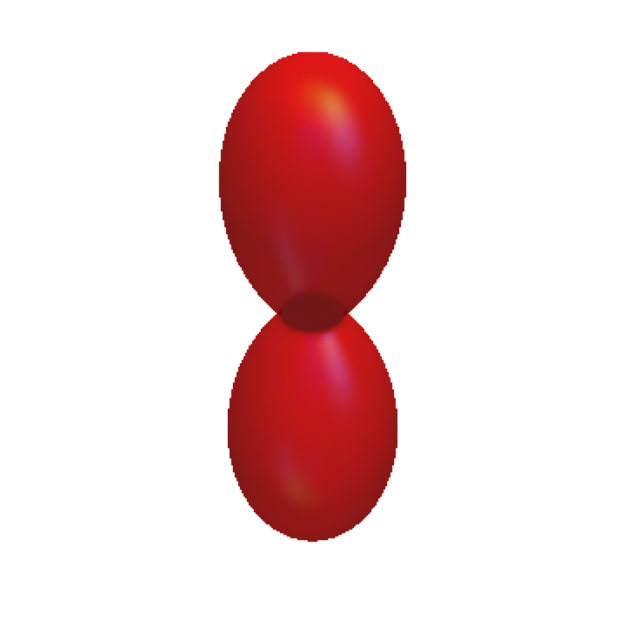}{0.15\textwidth}{Uniform field}
          \fig{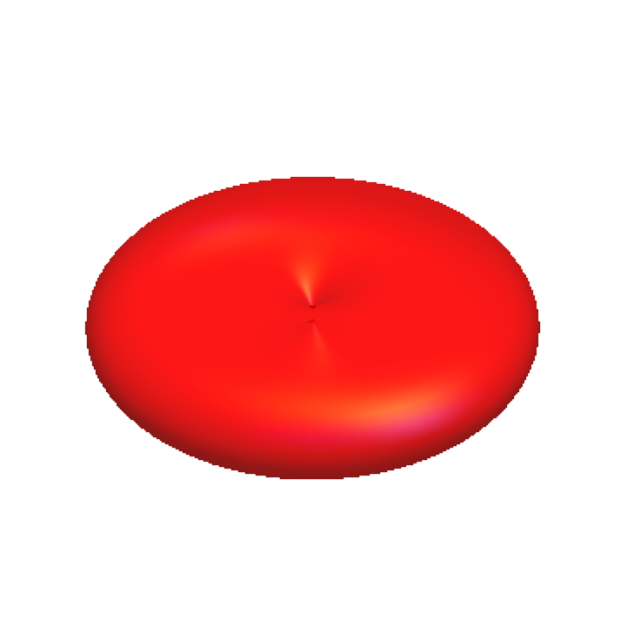}{0.15\textwidth}{Positive charge}
          \fig{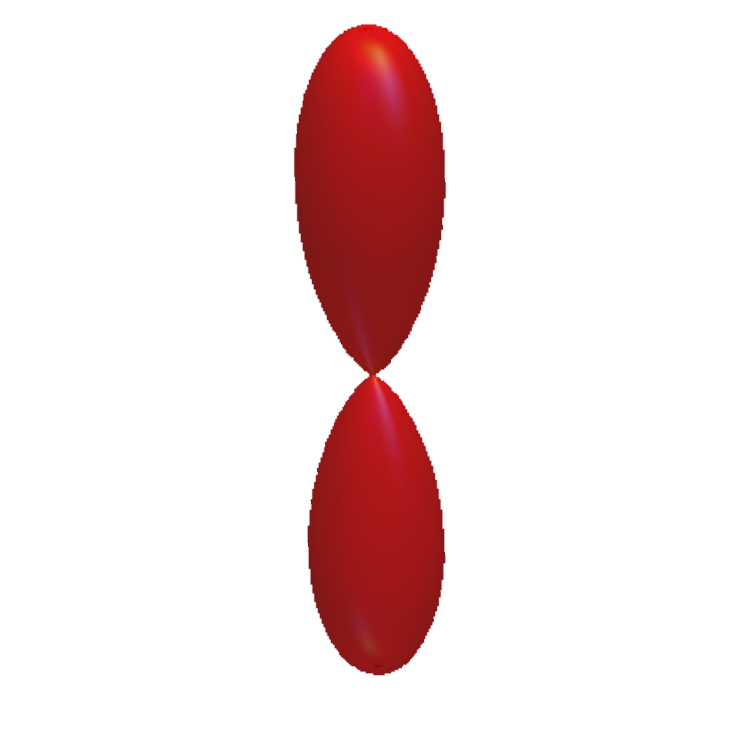}{0.15\textwidth}{Negative charge}}
\caption{Angular PDFs for the \emph{para}-\htwo\ rovibrational ground state in a field of strength $0.01$ (upper row), and $0.04$ atomic units (lower row). The $z$-axis (field direction) is vertically oriented in all cases. Left: a uniform field. Middle: a point-like positive charge. Right: a point-like negative charge.}
\label{fig:angularpdfs}
\medskip
\end{figure}

An important qualitative aspect of the solutions is that the $(2j+1)$-fold degeneracy of the unperturbed states is not completely lifted by the electrical perturbation (\ref{eq:electricalenergy}): for $m\ne0$ there remains a two-fold degeneracy associated with the sign of $m$. Consequently we will typically not make explicit mention of the $m<0$ states because their properties are identical to those of the corresponding $m>0$ states. Furthermore, the electrical perturbation does not mix states of different $m$ values so it is possible to restrict attention to $m\ge0$ at the outset. Doing so is beneficial because it reduces the size of the basis to only $1{,}861/1{,}828$ elements for the \emph{para-}/\emph{ortho-} sequences, effecting a factor of a few speed-up in the diagonalisation of the Hamiltonian. An additional benefit is that the radiative transition rates (\S4) are easier to interpret in the smaller basis --- because the perturbed eigenstates necessarily correspond to a well-defined value of $m$, rather than a linear combination of $\pm m$ as is the case with the full basis.

\subsection{The perturbed ground state}
To illustrate how an electric field affects the \htwo\ molecule we show its influence on the rovibrational ground state. The ground state is a useful point of reference because its properties in zero field are already familiar: an isotropic angular distribution of the internuclear axis, and a radial wavefunction that is approximately Gaussian.

The wavefunction of a perturbed state with eigenvector $\mathbf{a}$ can be written as 
\be
\Psi = \sum_{v,j} a_{v,j,m}\, \chi_{v,j}(r)\, Y_{j,m}(\theta,\phi),\label{eq:wavefunction}
\ee
and the probability, ${\rm d}P$, of the internuclear vector having length within ${\rm d}r$ and orientation within a solid angle ${\rm d}\Omega$ is ${\rm d}P=\Psi^*\Psi\,r^2\, {\rm d}r\, {\rm d}\Omega$. To obtain either the angular or radial probability density functions (PDFs) we expand $\Psi^*\Psi$ into a double sum and integrate over either the radial or the angular variables, respectively. Because the spherical harmonics are orthonormal the summation simplifies to
\be
{{{\rm d}P}\over{{\rm d}r}} = \sum_{v,j,v^\prime} a_{v,j,m}\,a_{v^\prime,j,m}^*\,{\cal R}_{v,j}(r)\,{\cal R}_{v^\prime,j}(r),\label{eq:radialprobabilitydensity}
\ee
for the radial PDF. (There is no comparable simplification for the angular PDF.) In practice it is convenient to construct approximate PDFs in which the smallest components of $\mathbf{a}$ are set to zero prior to evaluating the sum, and the results shown in this section were computed using only the largest 300 components.

\begin{figure}
\centering
\fig{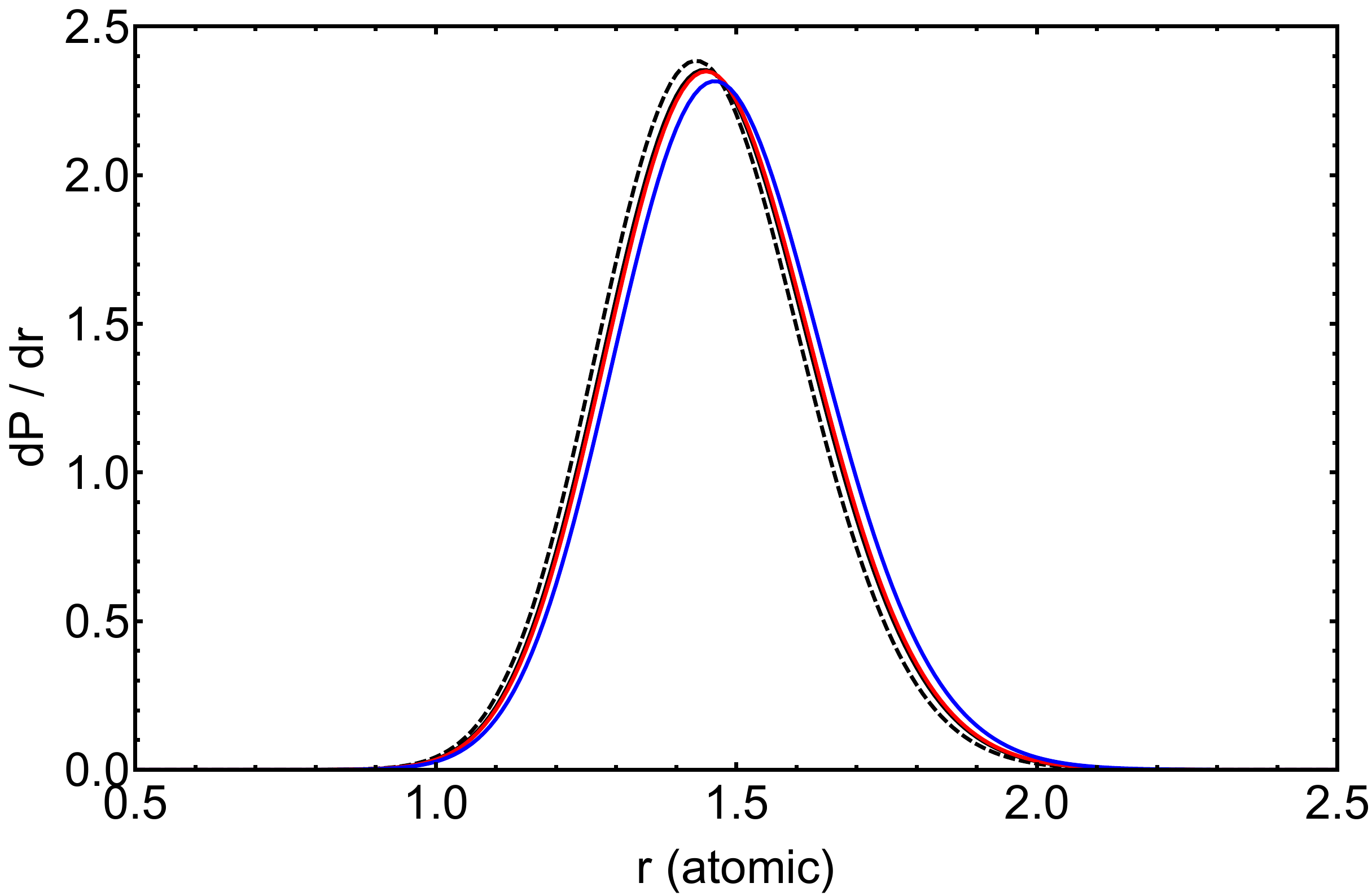}{0.45 \textwidth}{}
\vskip -0.5cm
\caption{Radial PDFs for the rovibrational ground state of \htwo, as follows: the field-free case, dashed black line; a uniform field, solid black line; positive charge, solid red line; and, negative charge, solid blue line. In the electrified cases the field strength is $0.04$ atomic units. The distributions are almost identical for a uniform field and for a positive charge. We do not show results for the weaker of the two fields ($0.01$ atomic units), because all three electrified curves lie on top of the zero-field distribution in that case.}
\label{fig:radialpdfs}
\medskip
\end{figure}

We have evaluated these PDFs for two different field strengths, $|{\mathbf F}|=0.01$ and $0.04$ atomic units, and for each of the three field configurations under study here: a uniform field; the field of a point-like charge $q=+e$; and, the field of a point-like charge $q=-e$. For each of the latter two cases the charge is located at separations of $R=1/\sqrt{|\mathbf{F}|}=10$ and $5$ atomic units ($R\simeq 5.13\,{\rm \AA}$ and $2.65\,{\rm \AA}$) from the centre-of-mass of the molecule.\footnote{The atomic unit of electric field strength corresponds to approximately $5.14\times10^{11}\;{\rm V\,m^{-1}}$. For a point charge, $q$, and separation, $R$, the radial electric field is simply $F=q/R^2$ when all quantities are expressed in atomic units. The fundamental charge $q=e$ is unity in atomic units, so $F=\pm 1/R^2$ for the positive and negative charges considered here.}

The effect of a uniform field is to align the molecule parallel to the field, as can be seen in the left-hand panels of figure~\ref{fig:angularpdfs}. This is as expected because the polarized state is energetically preferred, and the induced dipole moment is greatest when the field is oriented along the internuclear axis: \azz\ is significantly larger than \axx, and similarly for the hyperpolarizability, $\gamma$. The degree of alignment increases as the field strength increases.

When the molecule is near a point-like charge the perturbation is influenced by various electric field gradients, as well as the field strength, and the angular PDF changes accordingly. Some of the gradient-dependent perturbations -- specifically, those associated with $\Theta$, $\Phi$ and ${\rm B}$ (see Appendix B) -- change sign with the sign of the charge, leading to very different angular distributions in those two cases. At low fields the dominant perturbation is the $R^{-3}$ contribution from the permanent quadrupole, and the energetically preferred alignment for this interaction is parallel to the field if the charge is negative (top-right panel of figure~\ref{fig:angularpdfs}), and perpendicular if the charge is positive (top-middle panel of figure~\ref{fig:angularpdfs}). As the separation $R$ is decreased the other response tensors contribute, but the net effect is simply to enhance the weak-field alignment preferences --- as can be seen in lower-middle and lower-right panels of figure~\ref{fig:angularpdfs}. 

\begin{figure}
\centering
\fig{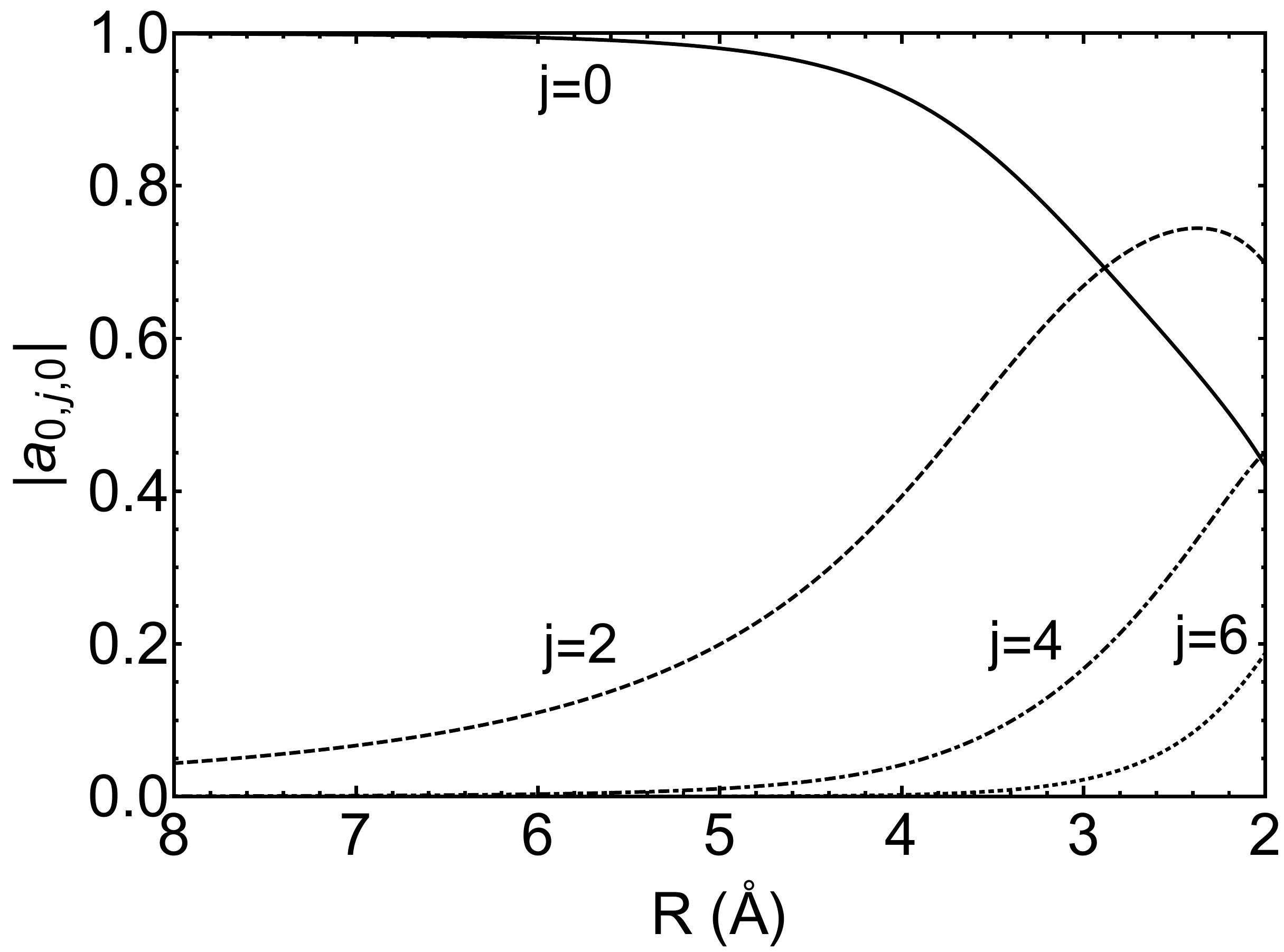}{0.45 \textwidth}{}
\vskip -0.5truecm
\caption{The moduli of the first few coefficients of the ground-state eigenvector, for a \emph{para-}\htwo\ molecule in the field of a point-like negative charge, plotted as a function of separation. The different lines show $|a_{0,j,0}|$ for: $j=0$ (solid line); $j=2$ (dashed line); $j=4$ (dot-dashed line); and, $j=6$ (dotted line).}
\label{fig:eigenvectorrotation}
\medskip
\end{figure}

The influence of the field on the radial distributions is not so pronounced because, as already noted, for the low-lying rovibrational states there is less mixing of vibrational eigenstates than rotational eigenstates. The radial PDFs derived from equation (\ref{eq:radialprobabilitydensity}) are presented in figure~\ref{fig:radialpdfs} (for field strength $0.04$), along with that of the unperturbed ground state, showing that the perturbed profiles are all very similar to each other. The main effect of the electrical perturbation is a slight shift of the centroid to larger radii. That can be understood by reference to the plots in Appendix A which show that, around the location of the ground state, all of the tensor components are increasing in magnitude towards larger radii. It follows that if the electrified state is energetically preferred then the ground state will shift to larger radii. And the perturbed states are indeed energetically preferred. Relative to the eigenvalue for the field-free ground state \citep[$-36{,}118.1\;{\rm cm^{-1}}$, ][]{komasa2011}, the electrified ground states in a field of strength $0.01$ ($0.04$) are located at: $-59.6\;{\rm cm^{-1}}$ ($-1{,}010.8\;{\rm cm^{-1}}$) for a uniform field; $-66.4\;{\rm cm^{-1}}$ ($-1{,}300.6\;{\rm cm^{-1}}$) for a positive charge; and, $-69.7\;{\rm cm^{-1}}$ ($-1{,}586.9\;{\rm cm^{-1}}$) for a negative charge.

\subsubsection{Eigenvector rotation with field strength}
Each of the eigenstates of the perturbed molecule can be described as a weighted sum of the unperturbed eigenstates,
\be
\ket{\Psi} = \sum_{v,j} a_{v,j,m}\,\ket{v,j,m},\label{eq:eigenvectorexpansion}
\ee
satisfying the eigenvector normalisation condition $|\mathbf{a}|^2=\sum_{v,j} |a_{v,j,m}|^2=1$. We can examine the behaviour of the coefficients $a_{v,j,m}$, for the ground state of the molecule, as the field strength increases; the evolution is shown in figure~\ref{fig:eigenvectorrotation} for the case of a \emph{para-}\htwo\ molecule in the field of a point-like negative charge.

At zero field the only non-zero coefficient of the ground state eigenvector is $a_{0,0,0}=1$, and as the field strength increases that eigenvector gradually moves away from \ket{0,0,0} (formally it is a rotation relative to the fixed basis vectors). In figure~\ref{fig:eigenvectorrotation}, therefore, we see $|a_{0,0,0}|$ declining as the field increases, while $|a_{0,2,0}|$ at first grows and then itself declines as the higher rotational states \ket{0,4,0} and \ket{0,6,0} start to contribute more strongly to the ground state.

\begin{figure}
\centering
\fig{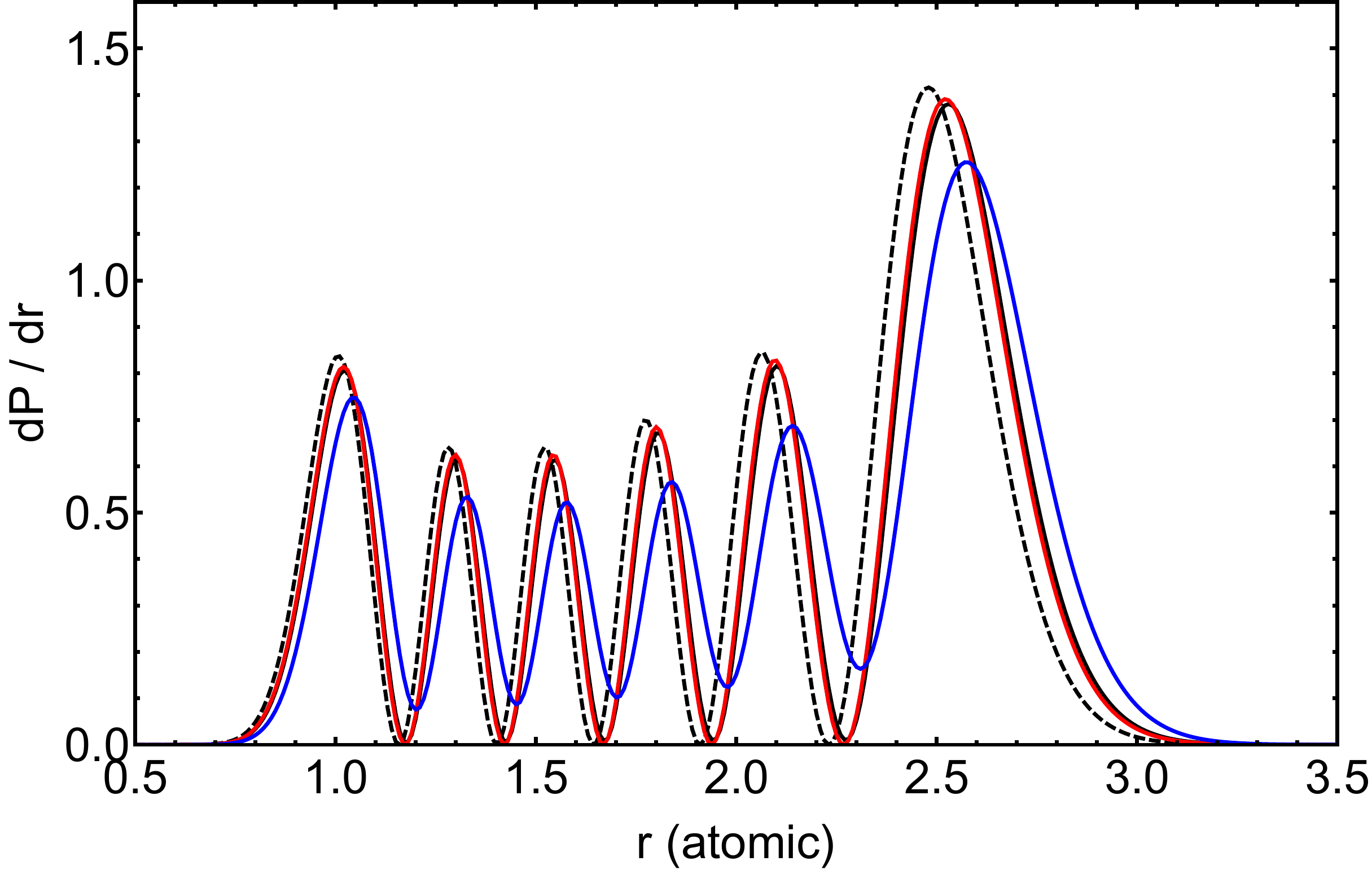}{0.45 \textwidth}{}
\vskip -0.5cm
\caption{As figure \ref{fig:radialpdfs}, but for the \ket{\hat{5},\hat{0},0} state of \htwo.}
\label{fig:excitedradialpdfs}
\medskip
\end{figure}

\subsection{Perturbed excited states}
The influence of an electric field on the excited rovibrational states  is qualitatively similar and we can therefore anticipate some basic trends in the excited states, as follows. First, as the total angular momentum of a rotational state increases so does the energy difference between adjacent rotational states. Consequently for a given strength of perturbation the degree of rotational mixing decreases as $j$ increases. The opposite is true for a vibrational sequence: the vibrational level separation shrinks as $v$ increases, so the degree of vibrational mixing goes up.

Of course the strength of the electrical perturbation is not actually the same along a rotational or vibrational sequence as the tensor components are functions of the internuclear separation (see Appendix A), and the expectation value of the internuclear separation differs from state to state. This change is also systematic in the sense that higher levels of rotational or vibrational excitation correspond to larger internuclear separations, and thus all of the tensor components increase in magnitude with increasing rotational or vibrational excitation. This effect is substantial even for the second-rank response tensors, with $\Theta$ and \axx\ reaching values up to about twice that of the ground state, and about three times for \azz\ (depending on the level of excitation). But it can be much larger for the fourth-rank response tensors --- particularly \gzzzz, which grows by about $20\times$, and \ezzzz\ which grows by about $40\times$ along the vibrational sequence. In combination with the decreasing vibrational quantum noted above this leads us to expect much greater vibrational mixing in the electrified states as the level of vibrational excitation increases --- a trend which has important consequences for the appearance of the ground state absorption spectra (see \S4).

The stronger mixing of the excited vibrational states is reflected in their radial PDFs, as illustrated with the \ket{\hat{5},\hat{0},0} state\footnote{Although $v$ and $j$ are not good quantum numbers for the electrified molecule, they are nevertheless useful for labelling the perturbed eigenstates and in that context we write them as $\hat{v}$ and $\hat{j}$ in order to distinguish \ket{\hat{v},\hat{j},m} from the unperturbed eigenstate \ket{v,j,m}. Labels are then assigned so that \ket{\hat{v},\hat{j},m} rotates smoothly to \ket{v,j,m} as the field is decreased smoothly to zero.} in figure \ref{fig:excitedradialpdfs} ($|{\mathbf F}|=0.04$ atomic units). In that figure we can see a much greater displacement of all the electrified states to larger separations (when compared to the displacement of the ground state seen in figure \ref{fig:radialpdfs}), and in the case of the negative charge (blue curve) the shape of the PDF is also changed substantially with local minima well above zero. The corresponding angular distributions are all qualitatively similar to examples already shown in figure \ref{fig:angularpdfs}, as follows: in the case of a positive charge the distribution is similar to the lower-middle panel of figure \ref{fig:angularpdfs}, whereas the uniform field and negative charge configurations are both similar to the lower-right panel of figure \ref{fig:angularpdfs}.

\subsection{Comparison with Poll \&\ Hunt (1985)}
The low-lying electrified states of \htwo\ have previously been characterized by \citet{pollhunt1985} for one of the field configurations studied here: \htwo\ in the field of a point-like charge. In particular \citet{pollhunt1985} calculated the energy levels of 16 low-lying states, for charges situated at various distances between $2\,{\rm\AA}$ and $6\,{\rm\AA}$ from the molecular centre-of-mass. Their calculations included only the second-rank response tensors, i.e. $\Theta$ and $\alpha$, whereas ours include also the fourth-rank response tensors, but they have addressed the same physical situation and it is appropriate to compare our results with theirs.

Quantitatively there are some differences between our results and those of \citet{pollhunt1985} even when we restrict our calculation to the second-rank response tensors in order to effect a like-for-like comparison. At the largest separations considered by \citet{pollhunt1985} our eigenvalues agree with theirs at the level of precision ($1\,{\rm cm}^{-1}$) that they were reported. Decreasing the separation to $3\,{\rm\AA}$ reveals some noticeable differences, of magnitude $\ga 2\,{\rm cm}^{-1}$. And as the separation between charge and molecule decreases further the differences increase rapidly, reaching $\sim 50\,{\rm cm}^{-1}$ for a positive charge and $\sim 200\,{\rm cm}^{-1}$ for a negative charge at $2\,{\rm\AA}$. The differences are systematic with all of our eigenvalues being more negative than the corresponding eigenvalues of \citet{pollhunt1985}.

\begin{figure}
\fig{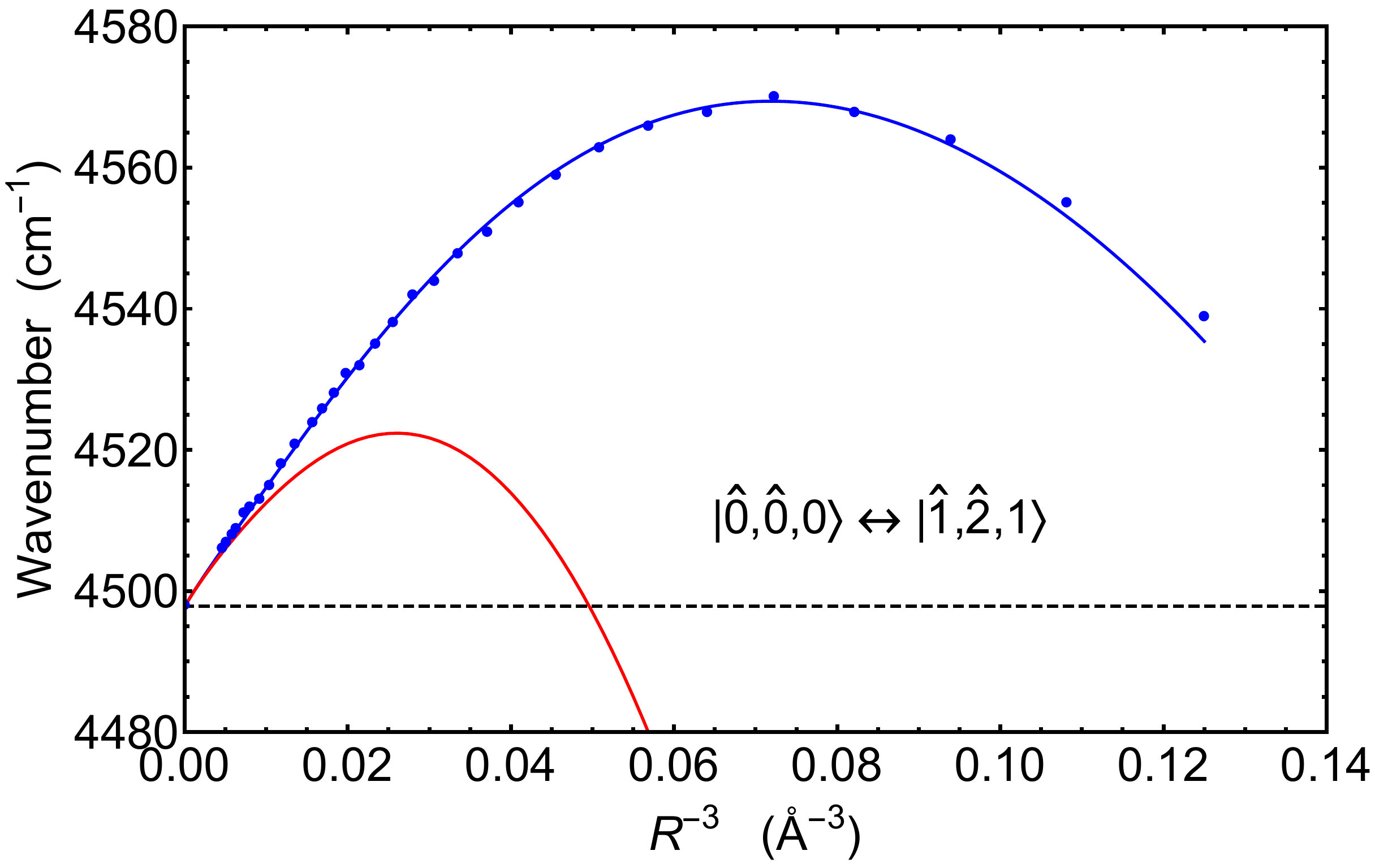}{0.45 \textwidth}{}
\vskip-0.5truecm
\caption{Transition wavenumbers for the fundamental vibrational line of \htwo\ situated in the field of a point-like charge $q=+e$, as a function of the (inverse-cube of the) separation, $R$, between the charge and the centre-of-mass of the molecule. Blue points show the results of \citet{pollhunt1985}; the blue line shows our like-for-like calculation ($2^{\rm nd}$ rank tensors only). The red line shows our full calculation, including the $4^{\rm th}$ rank response tensors. The dashed, black line shows the location of the transition in the field-free case. At large $R$ the quadrupole interaction dominates and the Stark shift is $\propto R^{-3}$ (see Appendix B).}
\label{fig:pollhuntfundamental}
\medskip
\end{figure}

Because the differences between the two sets of calculations are similar across the various eigenstates, the wavenumbers we obtain for transitions are in much better agreement than the eigenvalues themselves. Figure~\ref{fig:pollhuntfundamental} shows an illustrative example: the fundamental vibrational transition of \htwo, $|\hat{0},\hat{0},0\rangle\leftrightarrow|\hat{1},\hat{2},1\rangle$ in the field of a point-like positive charge. From the figure we can see that our like-for-like calculations agree very closely over most of the range plotted, with the differences being mainly due to quantisation in the \citet{pollhunt1985} values at the $1\,{\rm cm}^{-1}$ level. At the smallest charge separations the differences are larger but still quite modest at $\sim 4\,{\rm cm}^{-1}$. The good agreement we obtain with the results of \citet{pollhunt1985} gives confidence in our solutions.

Figure~\ref{fig:pollhuntfundamental} also shows the results of our full calculation for this transition, including both second- and fourth-rank response tensors. At the time of the \citet{pollhunt1985} analysis there was not enough information available on the fourth-rank response tensors to allow their influence to be gauged. But with the benefit of the recent work of \citet{miliordos2018}, and a huge increase in computational resources, it is straightforward to include all of the fourth-rank response tensors and we can now see that they are important. Their contribution is not completely negligible even at a separation of $6\;{\rm \AA}$, and at $2\;{\rm \AA}$ -- where the difference between blue and red curves reaches values $\sim 500\,{\rm cm}^{-1}$ -- they are an order of magnitude larger than the second-rank contribution to the Stark-shift of this spectral line. The rapid divergence of the fourth-rank calculation from the second-rank results, as the charge separation decreases, is understandable: the five extra tensors ($\Phi$, ${\rm E}$, ${\rm C}$, ${\rm B}$ and $\gamma$) contribute in proportion to steep powers of the separation, as can be seen in equations (B7), (B8) and (B9), so they are hundreds or thousands of times more important at $2\;{\rm \AA}$ than they are at $6\;{\rm \AA}$.

One point of practical importance in the case shown in figure~\ref{fig:pollhuntfundamental} is that the fourth-rank calculation (red curve) predicts a line shift that changes sign between large and small separations. Zero line shift occurs at a charge separation of approximately $2.7\;{\rm \AA}$, and for separations near that value this Stark-shifted line would be difficult to distinguish from a zero-field absorption.

\section{Absorption spectra}
Given the electrical energy perturbation, $\Delta E$, of equation (2), the electric dipole moment of the molecule is
\be
\mu_\alpha=-{{\partial \Delta E}\over{\partial F_\alpha}}\label{eq:dipolemomentgeneralexpression}
\ee
\citep{buckingham1967,bishop1990,miliordos2018}, which leads to
\begin{eqnarray}
\mu_\alpha = &
\alpha_{\alpha\beta} F_\beta 
+{1\over {15}}{\rm E}_{\alpha,\beta\gamma\delta} F^{\prime\prime}_{\beta\gamma\delta}\hskip2cm \nonumber \\
&+{1\over 3}{\rm B}_{\alpha,\beta,\gamma\delta}F_{\beta} F^\prime_{\gamma\delta}
\,+\,{1\over {6}}\gamma_{\alpha\beta\gamma\delta} F_{\beta} F_{\gamma} F_{\delta}.\hskip0.6cm\label{eq:dipolemomentspecificexpression}
\end{eqnarray}
As was the case with the energy perturbation itself, this particular form of the operator is not well suited to rapid evaluation of the corresponding matrix elements. Instead we make a spherical harmonic expansion, leading to the non-zero expansion coefficients listed in Appendix C. Like the Hamiltonian, the resulting matrices are highly sparse and can be conveniently stored in that form, being read in from file when they are needed. To determine the transition dipole matrix elements, i.e. \element{\Psi_1}{\mu_\alpha}{\Psi_2} between any two electrified states, \ket{\Psi_1} and \ket{\Psi_2}, simply requires a rotation of the matrix elements of $\mu$ from the unperturbed basis into the perturbed basis. Having evaluated the transition dipole matrix elements, the spontaneous radiative transition rate is given by \citep[e.g.][]{atkins2011} 
\be
A_{21}={4\over3} \kappa^3 |\element{\Psi_1}{\mu_\alpha}{\Psi_2}|^2,\label{eq:einsteina21}
\ee
where $\kappa\equiv 2\pi/\lambda$, and atomic units are used throughout. The corresponding Einstein $B_{12}$ absorption coefficient follows naturally, and the radiative lifetime of the upper state is simply the inverse of the sum of the spontaneous radiative decay rates to all lower-lying states, $\Gamma$:
\be
\Gamma_j=\sum_{i<j} A_{ji}.\label{eq:decayrate}
\ee

The spontaneous transition rate given by equation (\ref{eq:einsteina21}) obviously depends on the orientation of the vector $\mu_\alpha$, and thus on the polarisation state of the radiation under consideration. We will mostly ignore the polarisation properties of the transitions and deal only with the total spontaneous transition rate
\be
A_{21}=A_\parallel + A_\perp,
\ee
where the parallel component is given by
\be
|\element{\Psi_1}{\mu_\alpha}{\Psi_2}|^2\rightarrow  
|\element{\Psi_1}{\mu_z}{\Psi_2}|^2,
\ee
in equation (\ref{eq:einsteina21}), and the perpendicular component is given by
\be
|\element{\Psi_1}{\mu_\alpha}{\Psi_2}|^2\rightarrow  
|\element{\Psi_1}{\mu_x}{\Psi_2}|^2 + 
|\element{\Psi_1}{\mu_y}{\Psi_2}|^2.
\ee
Although we will not discuss the polarisation properties of the transitions, the breakdown into parallel and perpendicular transition rates is given in the numerical results in \S4.2 for readers who are interested in that aspect.

\begin{figure}
\fig{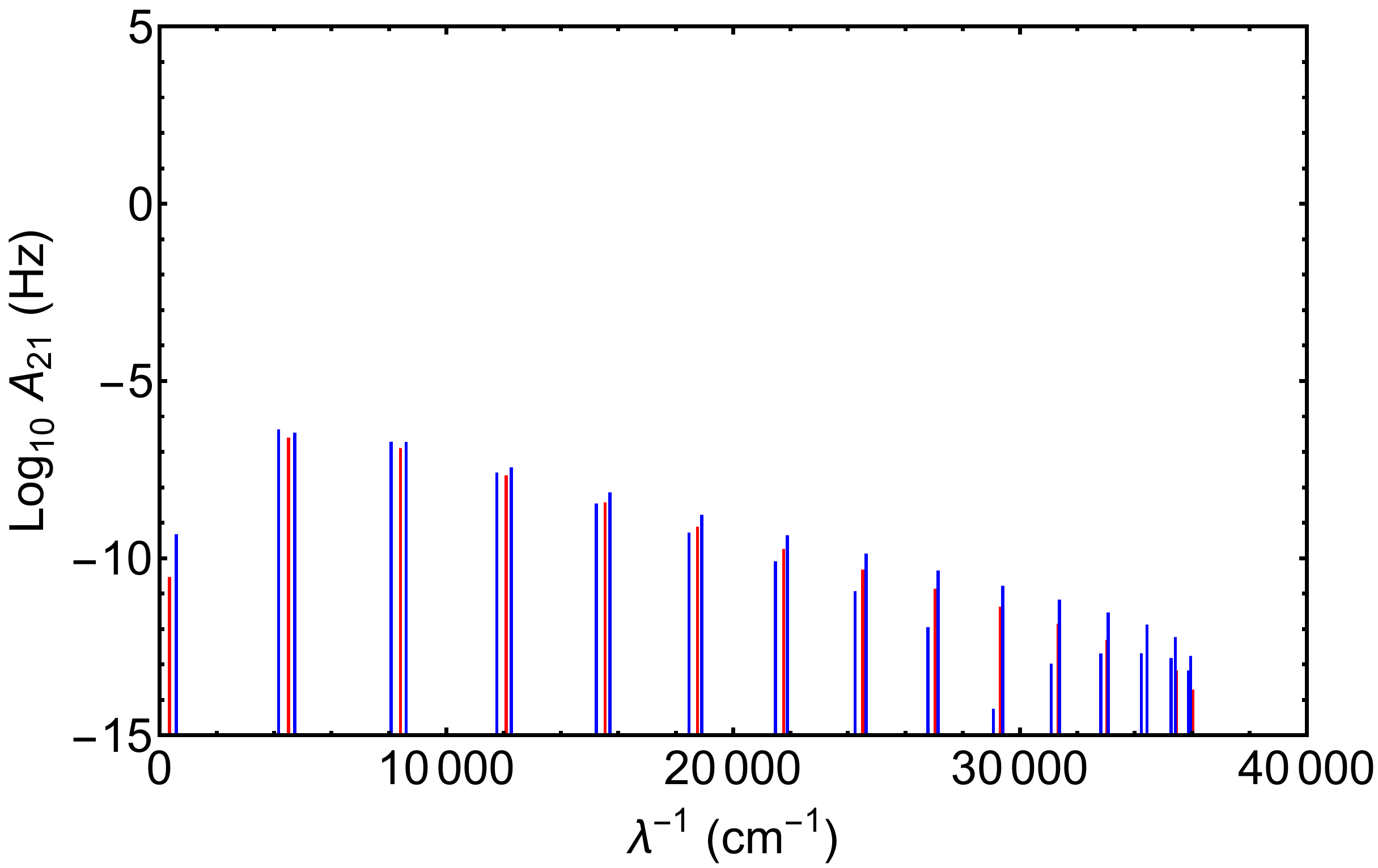}{0.45 \textwidth}{}
\vskip-0.5truecm
\caption{Transitions connecting to the rovibrational ground state for \emph{para-} (red) and \emph{ortho-} (blue) \htwo\ in the absence of an electric field. The line strengths are shown as Einstein $A_{21}$ spontaneous transition rates from the upper level via electric quadrupole and magnetic dipole radiation. All transition properties taken from \citet{roueff2019}.}
\label{fig:roueffspectrum}
\medskip
\end{figure}

\subsection{Zero-field absorption spectra}
Before considering the results of our calculations of the rovibrational absorption spectra of electrified \htwo\ it is helpful to have a clear picture of the rovibrational spectra of the field-free molecule. The main features of that spectrum are well known \citep[e.g.][]{herzberg1950,turner1977}, but a recent reassessment \citep[][]{pachuckikomasa2011,roueff2019} of the contribution of magnetic dipole radiation has led to some significant modifications to the theoretical spectrum. In figure~\ref{fig:roueffspectrum} we show the full spectra of ground-state rovibrational transitions, as computed by \citet{roueff2019}, for both \emph{ortho-} and \emph{para-} sequences of \htwo.

Although we are interested in absorptions from the ground state, it is simplest to display spectra as the spontaneous transition rates of the upper levels into the ground state -- i.e. the Einstein $A_{21}$ probabilities per unit time, as shown in figure~\ref{fig:roueffspectrum} -- rather than the $B_{12}$ coefficients, for example, whose values depend on the convention that is adopted for $B_{12}$. We will use the same format when displaying the spectra of the electrified molecule in the next section.

\begin{figure*}
\centering
\gridline{\fig{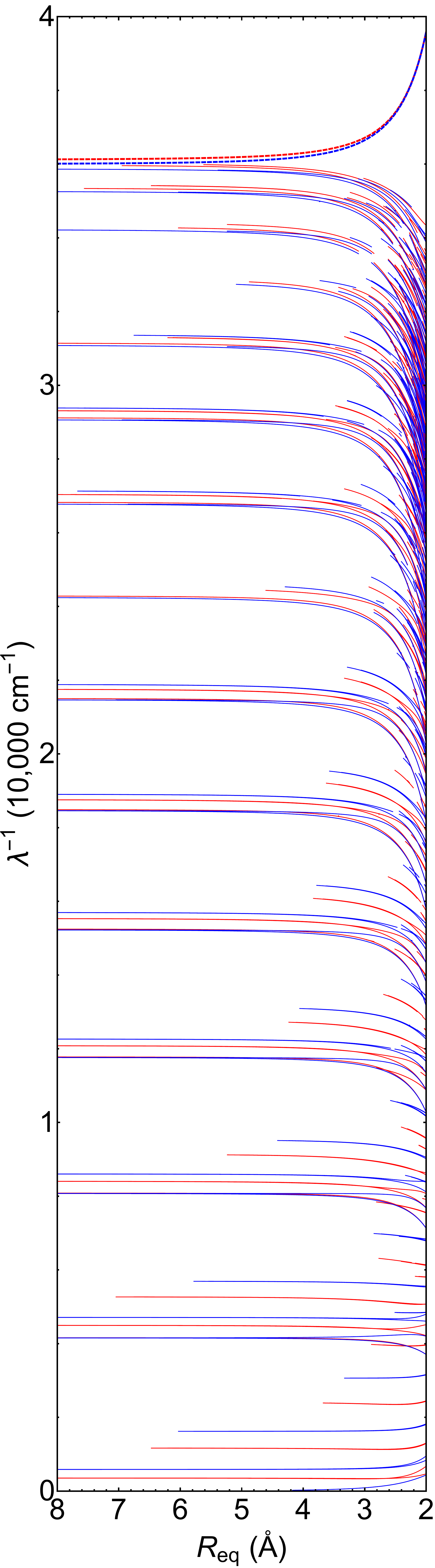}{0.29\textwidth}{}
          \fig{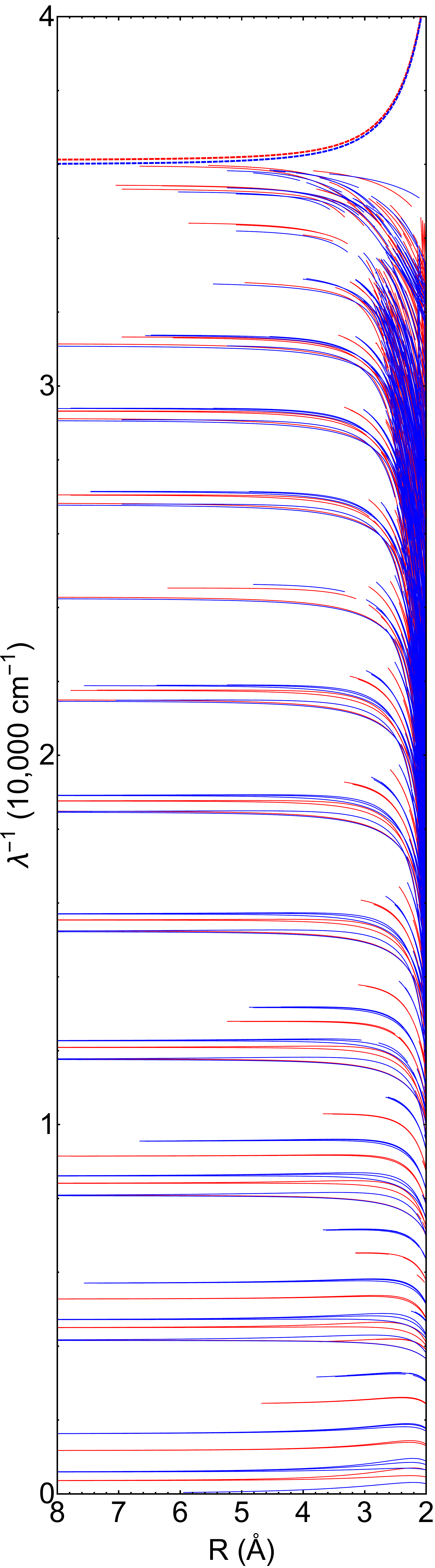}{0.29\textwidth}{}
          \fig{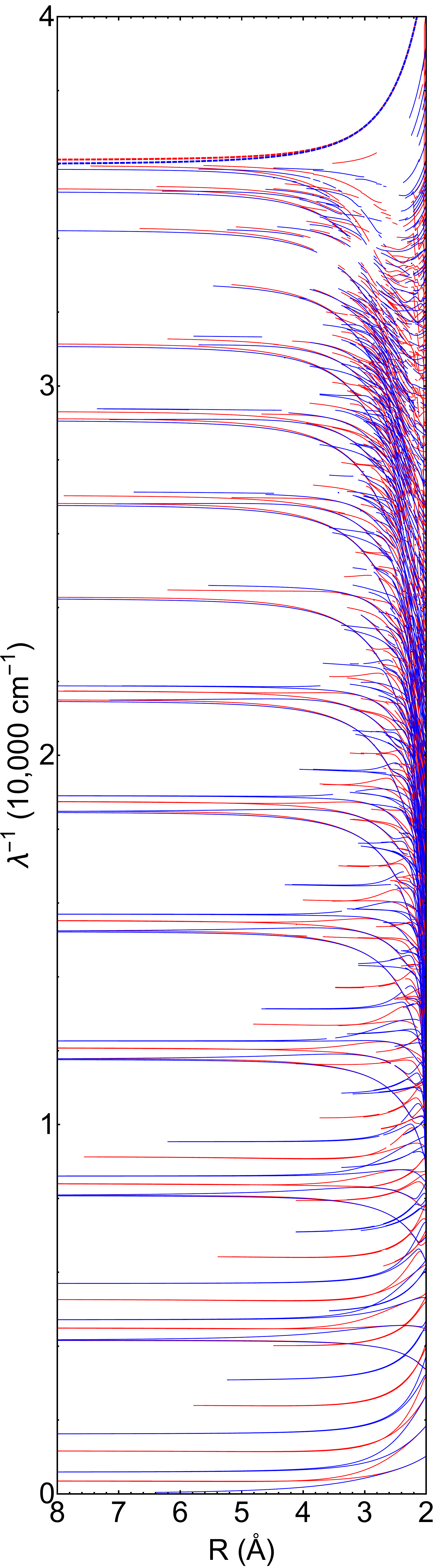}{0.29\textwidth}{}}
\vskip -0.7cm
\caption{Strong transitions ($A_{21}\ge 10^{-6}\;{\rm Hz}$) connecting to the ground state of electrified \htwo\  molecules in three different cases: a uniform field (left); the field of a point-like charge $q=+e$ (middle); and, the field of a point-like charge $q=-e$ (right). Transitions belonging to the \emph{para-} sequence are shown in red, those belonging to the \emph{ortho-} sequence are shown in blue; and the location of the continuum is shown with dashed lines of the appropriate colour. For each line the wavenumber is plotted (in ${\rm \mu m}^{-1}$) as a function of the separation, $R$, between the charge and the molecule, for the middle and right-hand panels. For the left-hand panel we use an analogous parameter, $R_{eq}$, to characterize the strength of the uniform electric field: in atomic units the field strength is $1/R_{eq}^2$.}
\label{fig:electrifiedspectra}
\medskip
\end{figure*}

Although magnetic dipole can be the principal radiation mechanism for some rovibrational transitions \citep[][]{roueff2019}, it is a small contribution (much less than 1\%) for all transitions connecting to the ground state of \emph{ortho-}\htwo, and zero for all transitions connecting to the ground state of \emph{para-}\htwo. Thus the qualitative appearance of figure~\ref{fig:roueffspectrum} is entirely familiar, consisting of the $S_v(0)$, $Q_v(1)$ and $S_v(1)$ lines, appearing in groups of the same upper-level vibrational quantum number, $v$, with $v=0,1,2\dots 14$. Anharmonicity in the internuclear potential is reflected in the decreasing spacing between line groups as the wavenumber is increased, and in the non-zero line strengths for transitions to states with $v\ge 2$. However, the level of anharmonicity is low enough that line strengths trend strongly downwards as $v$ increases.  

\subsection{Electrified absorption spectra}
In the case of electrified \htwo\ the molecular states naturally depend on the strength of the static field and we have therefore determined the energy eigensystem for various values of the field strength: 99 distinct values, equally spaced in the logarithm, for each of the field configurations under study. Rather than quoting the field strength itself it is more convenient to quote the separation, $R$, between the molecule and the point-like charge when considering that particular field configuration: we determined eigensystems for separations in the range $2\le R({\rm \AA}) \le 8$. In order to facilitate comparison between the different field configurations, in the case of a uniform field we construct an equivalent separation, $R_{eq}$, to characterize the electric field strength: $R_{eq}$ is the separation from a point-like, elementary charge that would yield the same field strength. In all cases we evaluated the electric dipole line strength for each transition connecting to the ground state of \emph{para-}\htwo\ and, separately, \emph{ortho-}\htwo, using equations (\ref{eq:dipolemomentspecificexpression}) and (\ref{eq:einsteina21}), with the results shown in  figure \ref{fig:electrifiedspectra}.

\vskip -1cm
\begin{deluxetable*}{cccccccc}
\tablecaption{Transitions of \htwo\ in a uniform field}
\tablehead{
\colhead{$R_{eq}$} & \colhead{$\hat{v}$} & \colhead{$\hat{j}$} & \colhead{$m$} & \colhead{$\lambda^{-1}$} & \colhead{$A_\parallel$} & \colhead{$A_\perp$} & \colhead{$\Gamma$}\\
\colhead{$({\rm \AA}$)} &  & & & \colhead{$({\rm cm^{-1}}$)} & \colhead{$({\rm Hz}$)} & \colhead{$({\rm Hz}$)} & \colhead{$({\rm Hz}$)}}
\startdata
  0.80000000E+01 & 0 & 0 & 0 & $-$0.36129451E+05 & 0 & 0 & 0\\
  0.80000000E+01 & 0 & 2 & 0 & 0.35349477E+03 & 0.64426370E$-$03 & 0 & 0.64426370E$-$03\\
  0.80000000E+01 & 0 & 2 & 1 & 0.35390320E+03 & 0 & 0.48325641E$-$03 & 0.48325641E$-$03\\
  0.80000000E+01 & 0 & 4 & 0 & 0.11678307E+04 & 0.78893398E$-$07 & 0 & 0.99479714E$-$02\\
  0.80000000E+01 & $\dots$ & $\dots$ & $\dots$ & $\dots$ & $\dots$ & $\dots$ & $\dots$\\
  0.80000000E+01 & 0 & 1 & 0 & $-$0.36012129E+05 & 0 & 0 & 0\\
  0.80000000E+01 & 0 & 1 & 1 & 0.17212478E+01 & 0 & 0.33195119E$-$10 & 0.33195119E$-$10\\
  0.80000000E+01 & 0 & 3 & 0 & 0.58730199E+03 & 0.22994695E$-$02 & 0 & 0.34442652E$-$02\\
  0.80000000E+01 & $\dots$ & $\dots$ & $\dots$ & $\dots$ & $\dots$ & $\dots$ & $\dots$\\
  0.78876298E+01 & 0 & 0 & 0 & $-$0.36130114E+05 & 0 & 0 & 0\\
  0.78876298E+01 & 0 & 2 & 0 & 0.35344409E+03 & 0.68172725E$-$03 & 0 & 0.68172725E$-$03\\
  0.78876298E+01 & 0 & 2 & 1 & 0.35387604E+03 & 0 & 0.51136096E$-$03 & 0.51136096E$-$03\\
  0.78876298E+01 & 0 & 4 & 0 & 0.11677748E+04 & 0.93529357E$-$07 & 0 & 0.10527520E$-$01\\
  $\dots$ & $\dots$ & $\dots$ & $\dots$ & $\dots$ & $\dots$ & $\dots$ & $\dots$
\enddata
\tablecomments{Transitions are listed in order of increasing wavenumber, at fixed $R_{eq}$, first for \emph{para-}\htwo\ and then for \emph{ortho-}\htwo; then $R_{eq}$ is decreased and the pattern repeats. There are 99 distinct values of $R_{eq}$, equally spaced in $\log R_{eq}$, covering the range $8\ge R_{eq}({\rm \AA})\ge 2$. For each combination of $R_{eq}$ and nuclear spin the ``wavenumber'' of the first line is actually the eigenvalue of the ground state (negative in all cases), with $\Gamma$ and both $A$-values set to zero. The complete table includes $\sim\,$3$\,\times 10^4$ lines and is available as an ascii text file.}
\end{deluxetable*}

\vskip -1.2cm
\begin{deluxetable*}{cccccccc}
\tablecaption{Transitions of \htwo\ in the field of a charge $q=+e$}
\tablehead{
\colhead{$R$} & \colhead{$\hat{v}$} & \colhead{$\hat{j}$} & \colhead{$m$} & \colhead{$\lambda^{-1}$} & \colhead{$A_\parallel$} & \colhead{$A_\perp$} & \colhead{$\Gamma$}\\
\colhead{$({\rm \AA}$)} &  & & & \colhead{$({\rm cm^{-1}}$)} & \colhead{$({\rm Hz}$)} & \colhead{$({\rm Hz}$)} & \colhead{$({\rm Hz}$)}}
\startdata
  0.80000000E+01 & 0 & 0 & 0 & $-$0.36130078E+05 & 0 & 0 & 0\\
  0.80000000E+01 & 0 & 2 & 1 & 0.35858861E+03 & 0 & 0.48218100E$-$03 & 0.48219198E$-$03\\
  0.80000000E+01 & 0 & 2 & 0 & 0.36301186E+03 & 0.68032214E$-$03 & 0 & 0.68032237E$-$03\\
  0.80000000E+01 & 0 & 4 & 1 & 0.11753055E+04 & 0 & 0.35392560E$-$05 & 0.94182994E$-$02\\
    $\dots$ & $\dots$ & $\dots$ & $\dots$ & $\dots$ & $\dots$ & $\dots$ & $\dots$
\enddata
\tablecomments{See comments for Table 1, but with $R_{eq}\rightarrow R$.}
\end{deluxetable*}

\vskip -0.9cm
\begin{deluxetable*}{cccccccc}
\tablecaption{Transitions of \htwo\ in the field of a charge $q=-e$}
\tablehead{
\colhead{$R$} & \colhead{$\hat{v}$} & \colhead{$\hat{j}$} & \colhead{$m$} & \colhead{$\lambda^{-1}$} & \colhead{$A_\parallel$} & \colhead{$A_\perp$} & \colhead{$\Gamma$}\\
\colhead{$({\rm \AA}$)} &  & & & \colhead{$({\rm cm^{-1}}$)} & \colhead{$({\rm Hz}$)} & \colhead{$({\rm Hz}$)} & \colhead{$({\rm Hz}$)}}
\startdata
  0.80000000E+01 & 0  & 0 & 0 &  $-$0.36130214E+05  &  0  &  0  &  0\\
  0.80000000E+01  &  0  & 2 & 0 &  0.34570893E+03  &  0.64714783E$-$03  &  0  &  0.64714783E$-$03\\
  0.80000000E+01  &  0  & 2 & 1 &  0.34997736E+03  &  0  &  0.46922604E$-$03  &  0.46922611E$-$03\\
  0.80000000E+01  &  0  & 4 & 0 &  0.11603530E+04  &  0.13245084E$-$04  &  0  &  0.10028004E$-$01\\
  $\dots$ & $\dots$ & $\dots$ & $\dots$ & $\dots$ & $\dots$ & $\dots$ & $\dots$
\enddata
\tablecomments{See comments for Table 1, but with $R_{eq}\rightarrow R$.}
\end{deluxetable*}

Because the electrified molecule has many absorption lines, only the strong ones are displayed in figure~\ref{fig:electrifiedspectra}; specifically we show only transitions with $A_{21}\ge 10^{-6}\;{\rm Hz}$ --- a limit that is higher than any of the corresponding transition rates of the field-free molecule (figure~\ref{fig:roueffspectrum}). Nevertheless it is clear that a large number of lines lie above that threshold ($\sim3\times 10^2$ in each of the \emph{para-} and \emph{ortho-} sequences in each of the three panels). Each of the three field configurations shows qualitatively similar spectral structure, as follows. In weak fields there is a small number of distinct lines, occurring in groups, whose wavenumbers are almost independent of the field strength and which lie close to the wavenumbers of the corresponding zero-field transitions (figure~\ref{fig:roueffspectrum}). In this regime the eigenstates of the electrified molecule are very similar in character to those of the field-free molecule, but the transitions are much stronger because of the electric dipole moment that is induced by the static field. As the field strength increases, new transitions appear in the plot where their $A_{21}$ values cross our chosen threshold; wavenumbers shift gradually, and degeneracies are lifted because different values of $m$ correspond to different expectation values of the electrical perturbation (\ref{eq:electricalenergy}). 

In all three panels of figure (\ref{fig:electrifiedspectra}) we can see examples of transitions moving to both higher and lower wavenumbers as the field strength increases. There are, however, systematic differences amongst these field configurations, with the uniform field case and the positive charge both showing a predominance of downward-trending wavenumbers whereas in the case of the negative charge the distribution between upward- and downward-trending lines is more even. A downward-trending line indicates that the upper level is moving closer to the ground-state as the field strength increases.

At the right-hand side of each of the panels in figure~\ref{fig:electrifiedspectra} the perturbation introduced by the static field is sufficiently strong that the states of the electrified molecule bear little resemblance to those of field-free \htwo, and at wavenumbers above $\sim 1\;{\rm \mu m^{-1}}$ the spectrum becomes a dense forest of strong lines. Indeed the spectra are so rich that it is difficult to represent all of the information in any plot which covers a range of different field strengths. In the next section (\S4.3) we will show an example for a specific choice of $R$ where the spectral structure can be seen in detail.

To allow readers to explore any aspect of these spectra that might be of interest, tables 1, 2 and 3 present our results for wavenumber, transition probability and natural linewidth, $\Gamma$ from equation (\ref{eq:decayrate}), for each transition, for each of the three field configurations, and each of the 99 field strengths considered. These tables are more comprehensive than the spectra of figure~\ref{fig:electrifiedspectra} in two respects: all transitions with probabilities $A_{21}\ge 10^{-12}\;{\rm Hz}$ are given; and, the transition rates for parallel ($A_\parallel$) and perpendicular polarizations ($A_\perp$) are given separately.

All of the transitions are fully linearly polarized, with our calculated numerical values of the transition rates for one mode being typically twenty orders of magnitude larger than the other; the actual numerical value of the  transition probability obtained for the weaker mode is therefore uninteresting and has been set to zero in our tabulations. We note that ``parallel'' transitions ($A_\perp=0$) are exclusively associated with transitions between states having the same value of $m$, whereas ``perpendicular'' transitions ($A_\parallel=0$) involve a change $\Delta m =\pm 1$. That division of polarisations is manifest in tables 1-3, and is expected on the basis of the spherical harmonic expansion coefficients of the dipole moment operator given in Appendix C.

Because of the two-fold degeneracy in the perturbed levels, corresponding to the two possible signs of $m$ (when $|m|>0$), we have not reported on $m<0$ states in tables 1-3. However, readers must remember those states, and correctly account for them, when making use of these tables. For example: the third line in each table gives $A_{21}=A_\perp$ for the transition $\ket{\hat{0},\hat{2},1}\rightarrow\ket{\hat{0},\hat{0},0}$, which implies that there is a corresponding transition $\ket{\hat{0},\hat{2},-1}\rightarrow\ket{\hat{0},\hat{0},0}$ with the same wavenumber and $A_{21}$, but which does not appear in the table. The strength of the absorption line at that wavenumber is the combined strength of the two transitions $\ket{\hat{0},\hat{0},0}\rightarrow\ket{\hat{0},\hat{2},\pm 1}$.

\subsubsection{Eigenvector rotation and state labels}
As noted earlier, $v$ and $j$ are not good quantum numbers for the electrified molecule, so the values $\hat{v}$ and $\hat{j}$ reported in tables 1, 2 and 3 are merely convenient labels. Even so there is no rigorously correct way of assigning them; our labels have been determined in the following way. In combination the three quantum numbers $\{v,j,m\}$ uniquely define an eigenstate in zero field. For the weakest field that we considered ($R,R_{eq}=8\;{\rm \AA}$), the character of the eigenstates is similar to those of the zero-field case, so we simply form the inner product of each of the electrified eigenstates with each of the zero-field eigenstates and transfer the labels from the old to the new according to the magnitudes of those inner products. To ensure the mapping is one-to-one the transfer is done sequentially, starting from the electrified ground state and working up the eigenvalue spectrum, and removing each label from the available pool once it has been assigned. This procedure is then repeated, transferring labels from the eigenstates of the weakest field we considered to those of the next-weakest, and so on.

If the field is so weak that the perturbed eigenstates are all very close to the unperturbed eigenstates then there is very little ambiguity in the assignment of labels and the procedure described above should yield useful results. However, once the field is strong enough that there is little direct correspondence between the perturbed and unperturbed states then the choice of labels becomes less clear-cut. Figure \ref{fig:eigenvectorrotation} shows that we are certainly in that regime for the strongest fields under consideration, because even the ground state $\ket{\hat{0},\hat{0},0}$ has rotated a long way from $\ket{0,0,0}$ by the time we reach separations $R\simeq 2\,{\rm \AA}$. Moreover we caution that our coverage of $R$ and $R_{eq}$ -- i.e. 99 points, equally spaced in the logarithm -- might not adequately sample the rotation of the excited state eigenvectors, making it impossible to reliably transfer labels from one field strength to the next.

\begin{figure}
\fig{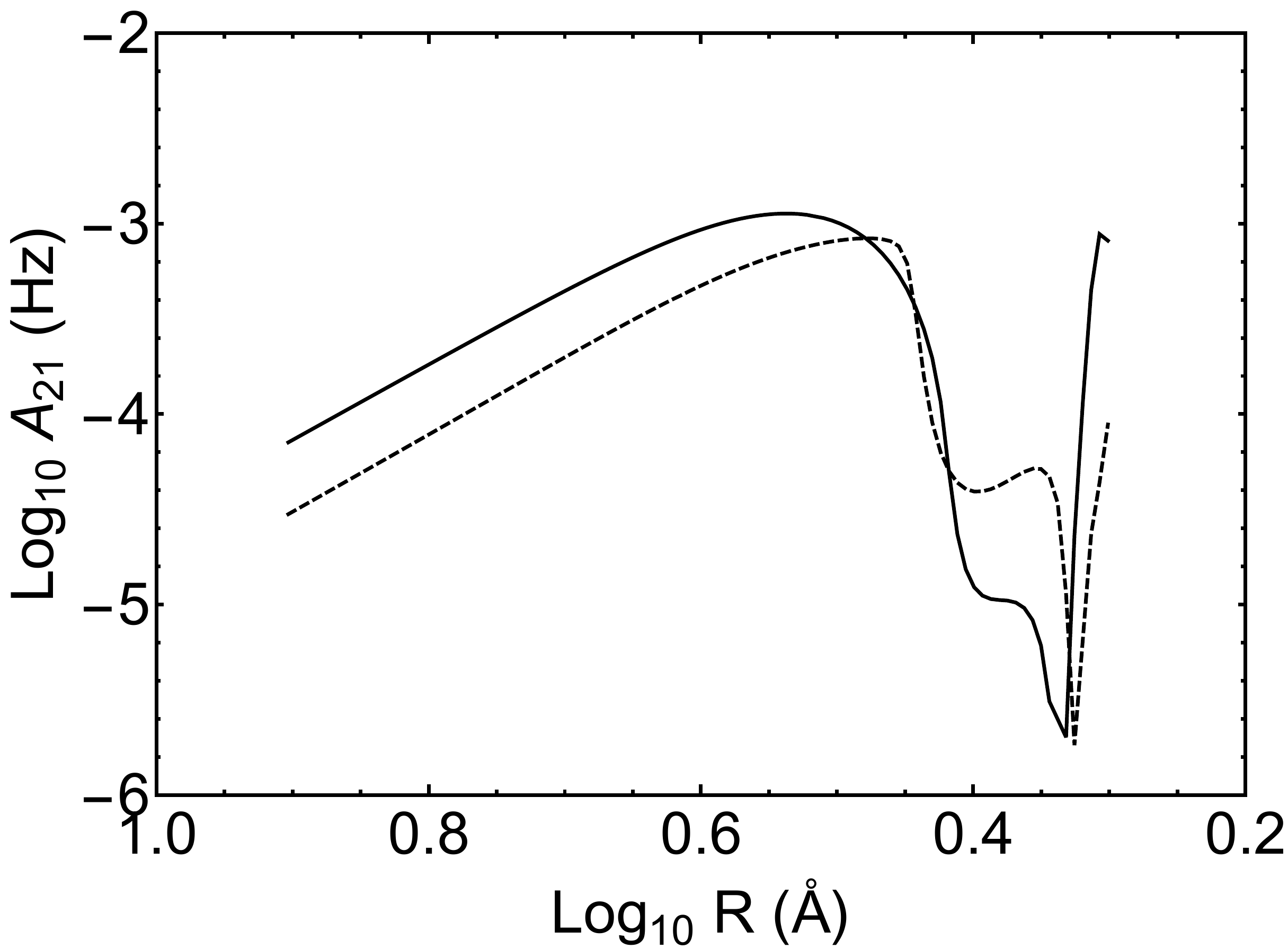}{0.45 \textwidth}{}
\vskip-0.5truecm
\caption{Variation of line strengths with separation from a point-like charge $q=-e$, for transitions connecting the ground state \ket{\hat{0},\hat{0},0} to \ket{\hat{3},\hat{2},1} (solid line), and \ket{\hat{4},\hat{2},1} (dashed line). At large separations the line strength varies approximately as $R^{-4}$, whereas at small separations the behaviour is complicated, and substantially different between the two lines.}
\label{fig:linestrengths}
\medskip
\end{figure}

\subsubsection{The dependence of line strength on field strength}
Some simple properties of the field-induced line strengths can be anticipated, according to the following argument. In weak-fields the eigenstates can be approximated by the unperturbed eigenstates, so we expect the strongest transitions to arise between two states that have a non-zero value of the electric dipole matrix element in the angular momentum basis, i.e. \element{j,m}{\mu_\alpha}{j^\prime,m^\prime}. Now the values of those matrix elements scale approximately linearly with the electric field strength (the first two terms on the right-hand side of equation \ref{eq:dipolemomentspecificexpression}), leading to line strengths that scale approximately as $R^{-4}$ (or $R_{eq}^{-4}$). That expectation is borne out in practice, as can be seen in figure \ref{fig:linestrengths} where we show two examples for the case of a molecule in the field of a point-like charge $q=-e$. We also note that the transition wavenumbers are approximately constant in weak fields (see figure \ref{fig:electrifiedspectra}), so in that regime the shape of the whole absorption spectrum is fixed and only the overall normalisation varies with the field strength.

In contrast to the very simple situation in weak fields, figure \ref{fig:linestrengths} also demonstrates complicated behaviour in the line strengths at small separations. Nor is the form of the variation the same for the two lines, and in fact the line ratio crosses unity several times at separations below about $3\;{\rm \AA}$. The reasons behind the complicated behaviour are easy to understand by reference to figure \ref{fig:eigenvectorrotation}: in strong fields the eigenvectors of both the ground state and the excited state are changing rapidly, relative to the unperturbed basis, as the field strength increases. That means that the contributions made by the individual dipole matrix elements \element{v,j,m}{\mu_\alpha}{v^\prime,j^\prime,m^\prime} (i.e. in the unperturbed basis) to the transition dipole \element{\hat{v}_1,\hat{j}_1,m_1}{\mu_\alpha}{\hat{v}_2,\hat{j}_2,m_2} must also vary rapidly as the separation changes. Remembering that most of the matrix elements  \element{j,m}{\mu_\alpha}{j^\prime,m^\prime} are zero (see Appendix C and Appendix B), it is no surprise that there are large, non-monotonic variations in line strength at small values of the separation $R$.

\subsubsection{A dearth of bound states near the continuum}
In figure~\ref{fig:electrifiedspectra} the continuum levels for \emph{para-}/\emph{ortho-}\htwo\ are shown as red/blue dashed lines near the top of each panel. In all cases we see that the continuum level moves to higher wavenumbers as the field strength increases, simply reflecting the fact that the ground state becomes more tightly bound with increasing field strength. What is more interesting is that a gap opens up between the highest eigenvalues and the rovibrational continuum, and for strong fields that gap is large compared to the spacing between the eigenvalues. That is true of all three field configurations, but the gap is particularly large for a uniform field and for the field of a positive charge.

It seems likely that this gap is simply an artifact of our model, arising from incompleteness in the basis --- a possibility that we advertised at the start of \S2. The point is that all of the states become more tightly bound when the molecule is situated in an electric field. Presumably, then, states that are unbound in the case of the unperturbed molecule may become bound in the presence of an electric field, and would thus fill in the gap. We return to this issue in \S6.1.

\subsubsection{A very low frequency transition}
In the field-free molecule the longest wavelength line is the \emph{para-}\htwo\ line $S_0(0)$ at $\lambda\simeq 28\;{\rm \mu m}$, so given that the Stark shifts are small in weak electric fields we might also expect this to be the longest wavelength line of the electrified molecule. Figure~\ref{fig:electrifiedspectra} shows that it is not: there is a transition of \emph{ortho-}\htwo\ with a wavelength that, in the weak field case, can be \emph{much} longer. Clearly this is not the familiar $S_0(1)$ transition -- which is the (blue) line seen, as expected, just above the (red) $S_0(0)$ line in figure~\ref{fig:electrifiedspectra} -- rather it is a ``pure orientational'' transition, in which only the $m$ value changes: $\ket{\hat{0},\hat{1},0}\leftrightarrow\ket{\hat{0},\hat{1},\pm 1}$. (Note that in the case of a uniform field, and also in the case of a negative point-like charge, the ground state of the \emph{ortho-} sequence has $m=0$ and the first excited state is the doublet $m=\pm 1$, whereas the reverse is true for the case of a positive charge.) Table 1 shows that in the case of a uniform field with $R_{eq}=8\,{\rm \AA}$ the wavenumber of this transition is only $\lambda^{-1}\simeq 1.72\;{\rm cm^{-1}}$; in this instance, however, it is a weak line with $A_{21}\simeq 3\times 10^{-11}\,{\rm Hz}$. In each of the three configurations, this line increases in strength and shifts to higher wavenumbers as the field strength increases. 

\bigskip
\subsection{A model spectrum appropriate to solvated \hminus}
In the previous section we presented results spanning a range of field strengths and different field configurations; now we focus on one particular circumstance, allowing us to present the calculated spectrum in more detail. The case we consider here is \htwo\ in the field of a point-like negative charge, at a separation of $R=2.7811\;{\rm \AA}$ ($5.2555$ atomic units). This choice was motivated by the results of \citet{wangandrews2004}, who concluded that the \hminus\ ion in solvation in condensed \htwo\ induces a strong line at $3{,}972\;{\rm cm^{-1}}$ in the ligand molecules; our quoted separation, $R$, was chosen so as to reproduce the wavenumber of that line in our model \emph{para-}\htwo\ sequence. This value of the separation should be interpreted as the radius of the first solvation shell of \hminus\ in condensed \htwo.

\begin{figure}
\fig{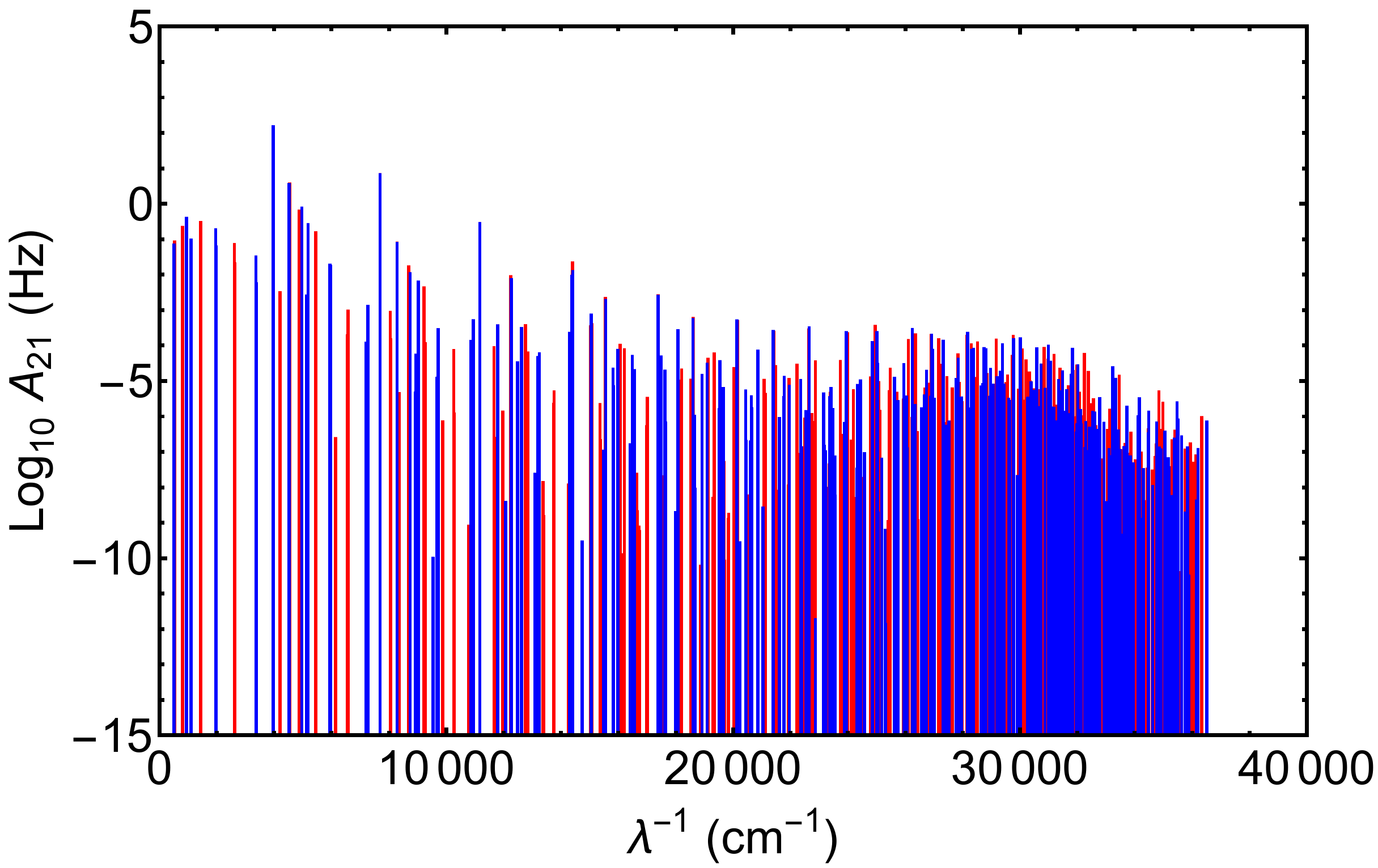}{0.45 \textwidth}{}
\vskip-0.5truecm
\caption{Transitions connecting to the rovibrational ground state for \emph{para-} (red) and \emph{ortho-} (blue) \htwo\ at a distance of $R=5.2555$ (atomic) from a point-like negative charge. The line strengths are shown as Einstein $A_{21}$ spontaneous transition rates from the upper level. Note that the scale on both axes is the same as figure~\ref{fig:roueffspectrum}.}
\label{fig:hminusspectrum}
\medskip
\end{figure}

The full spectrum obtained from our calculation is shown in figure \ref{fig:hminusspectrum}; the line mentioned above is the strongest \emph{para-}\htwo\ line in the plot. In fact our model actually produces two lines with almost identical characteristics: $\ket{\hat{0},\hat{0},0}\leftrightarrow\ket{\hat{1},\hat{0},0}$, with $\lambda^{-1}=3{,}972.0\;{\rm cm^{-1}}$ and $A_{21}=163\;{\rm Hz}$; and, $\ket{\hat{0},\hat{1},0}\leftrightarrow\ket{\hat{1},\hat{1},0}$, with $\lambda^{-1}=3{,}971.2\;{\rm cm^{-1}}$ and $A_{21}=164\;{\rm Hz}$. Because their properties are so similar, they are superimposed in figure \ref{fig:hminusspectrum} and only one can be discerned.

The spectrum in figure \ref{fig:hminusspectrum} is rendered on exactly the same scale as that of the zero-field spectrum shown in figure \ref{fig:roueffspectrum}, demonstrating very clearly that the lines of the electrified molecule are orders of magnitude stronger and more numerous than those of the field-free molecule. In common with the zero-field spectrum, figure \ref{fig:hminusspectrum} does exhibit a strong decline in line strengths from the fundamental vibrational transition to the overtone lines. Unlike the zero-field case, however, the decline is arrested at about the fifth overtone, with little obvious change in character over the range from $20{,}000$ to $30{,}000\;{\rm cm^{-1}}$.

After the fundamental vibrational lines, the next-strongest transitions seen in figure \ref{fig:hminusspectrum} are the overtone lines $\ket{\hat{0},\hat{0},0}\leftrightarrow\ket{\hat{2},\hat{0},0}$ and $\ket{\hat{0},\hat{1},0}\leftrightarrow\ket{\hat{2},\hat{1},0}$ at $\lambda^{-1}\simeq 7{,}692.3\;{\rm cm^{-1}}$ and $\lambda^{-1}\simeq 7{,}691.2\;{\rm cm^{-1}}$, respectively (both with $A_{21}\simeq 7.3\;{\rm Hz}$); again, only one of the pair can be discerned by eye. To the author's knowledge there has been no published report of a strong line near this wavenumber in studies of irradiated solid hydrogen crystals, even though some studies covered this spectral region \citep[e.g.][]{chan2000}. That is perhaps unsurprising given that the overtone is expected to have less than 5\% of the intensity of the fundamental, but it does mean that our spectral model lacks an  independent validation.

\section{Electrified \htwo\ in the ISM}
In this section we turn our attention to circumstances where the calculations of \S\S3,4 are relevant to astronomical observations --- i.e. manifestations of electrified \htwo\ in the ISM.

\subsection{Condensed environments}
We identify two main circumstances where \htwo\ could be immersed in a static electric field: a single ion with a small number of \htwo\ ligands, which we will refer to as a nanocluster; and, dust crystals of solid \htwo\ permeated by electric fields from embedded ions. Broadly speaking these cases represent the conventional and the unconventional faces, respectively, of condensed \htwo. Whereas nanoclusters are likely to form under conditions that are \emph{known} to exist in the ISM \citep[e.g.][]{duley1996}, if our Galaxy contains molecular gas that is close to the \htwo\ saturation curve then, as discussed in the Introduction, it would likely have escaped notice.

\subsubsection{Nanoclusters}
Nanoclusters may form whenever ionic species are present within molecular gas. In practice that means essentially all molecular gas clouds, because there is always some flux of radiation that is energetic enough to penetrate the cloud and cause ionisation therein; cosmic-rays are the usual culprits. The resulting chemistry may be complex and interesting, but for our purposes the identity of the ions is less important than the fact that they are able to nucleate the condensation of \htwo. Rates of formation and destruction and the resulting equilibrium abundances, for some types of \htwo\ nanoclusters, were considered by \citet{duley1996}.

Of all the possible astrophysical manifestations of electrified \htwo, the closest correspondence to our model calculations is the case of a ``nanocluster'' of one ion and one ligand \htwo\ molecule. Even in this case, however, our approach is only an approximation. First there is the interaction between the electrons in the molecule and those in the ion, which may contribute positively (repulsion) or negatively (bonding) to the total energy; this interaction is not captured by the description we are using, which is purely electrostatic in nature. Secondly, even our description of the electrostatics is an approximation because the ions are not simply point-like charges. Take the \hminus\ ion, for example (\S4.3): in isolation it is, of course, a spherically symmetric system; but it also has very high polarisabilities \citep[][]{pipin1992}, and under the influence of the electric field from a nearby, polarized \htwo\ molecule it can be expected to develop significant multipolar structure in its own charge distribution. All this is not to say that the results of \S4 are not useful, but that one should not expect a precise correspondence between calculated and observed spectra.

If the nanocluster consists of one ion and several ligand \htwo\ molecules then, of course, the system is not so well approximated by our calculations. Each of the ligands has induced multipole moments (dipole, quadrupole, etc.), in addition to the permanent quadrupole, hexadecapole etc., and these neighbouring molecules contribute to the electrical potential structure in the vicinity of the molecule under consideration. Unfortunately, calculating the multipole moments is a difficult task because they depend on the electric fields due to all neighbouring molecules, which in turn depend on the multipole moments themselves.

We have not attempted to quantify the mutual couplings just described, but there is a growing literature on \emph{ab initio} structural models of molecular hydrogen nanoclusters and such models could be used to assess the electrical environments in which the \htwo\ moieties reside. For nanoclusters of the type ${\rm H}^-{\rm (H_2)}_n$ structures have been determined using a variety of quantum chemical methods \citep[e.g.][]{huang2011,calvo2018,mohammadi2020}. We note that \citet{calvo2018} obtained results in accord with the mass spectroscopy of anion clusters undertaken by \citet{renzler2016}, who demonstrated the existence of ``magic numbers'' $n=12$, $n=32$ and $n=44$, consistent with icosahedral solvation shells.

Similarly, mass spectroscopy of cation clusters by \citet{jaksch2008} yielded results consistent with icosahedral solvation shells around the \hsix\ moiety. Unfortunately there have been no theoretical investigations confirming that interpretation --- the largest such nanocluster studied to date includes only four ligand \htwo\ molecules \citep[][]{kurosaki1998}. By contrast icosahedral solvation does not seem to be favoured in the case of the ${\rm H}_3^+$ moiety, with the data of \citet{jaksch2008} apparently indicating a preference for three or six ligand \htwo\ molecules. The latter preference was evident already in the early work of \citet{clampitt1969}. Detailed theoretical modelling of these small clusters has also been undertaken \citep[e.g.][]{barbatti2000,barbatti2001}.

Although the hydrogenic ions are expected to be important, because hydrogen is the most abundant element, there are  many possibilities for forming nanoclusters based on other ionic species --- see \citet{bernstein2013}, for example. The most important non-hydrogenic species are presumably those made up of elements with a high abundance, e.g. ${\rm OH}^-$, and, in the case of cations, a low ionisation potential for the parent, such as ${\rm Li}^+$. 

\subsubsection{Crystals of solid \htwo}
There is no clear limit to the number of ligand molecules that can be present before we are obliged to abandon the term ``nanocluster'' and use instead the name ``crystal'' or ``dust particle'', but a natural division arises where the assembly has more than one ion. As discussed in the Introduction, a dust crystal of \htwo\ in the ISM may have both positive ions and negative ions within the matrix, along with a two-dimensional electron gas trapped on the crystal surface. To some extent, then, we can think of each crystal of solid \htwo\ in the ISM as being an assembly of nanoclusters of various types. There are, however, some fine distinctions worth noting, as follows.

First it is not just the microscopic picture that is relevant. What we might measure in an astronomical context is also dictated, in part, by the overall structure of the condensate --- i.e. the size and shape of the solid particles. One can imagine a wide variety of possible forms, just as there are many different forms for snowflakes of water ice that are encountered on Earth --- dendritic, or plate-like, or columnar for example \citep[e.g.][]{libbrecht2017}. At present, however, we have no useful guidance on these aspects of the astrophysical problem of hydrogen ices, so it seems appropriate to set them aside for now.

Secondly, in a dust crystal it is possible that the bulk of the molecules are situated in a field that is approximately uniform, whereas this field configuration cannot arise in a nanocluster.

Thirdly, if a large number of nanoclusters are assembled into a dust particle then a huge number of configurations is possible, and a continuum description of the electrical environment would certainly be required. This is true even if we consider the uniform field configuration noted immediately above, because the field strength will be different in different particles.

\subsubsection{Macroscopic bodies of solid \htwo}
This section would not be complete without mention of the possible existence of macroscopic bodies of solid hydrogen. That is not a new idea \citep[e.g.][]{white1996}, but the possibility that we might have actually observed one -- in the form of the interstellar object `Oumuamua -- is a fascinating recent development \citep[][]{fuglistaler2018,seligman2020,levine2021}. Precisely because of their large dimensions, such objects will not contribute significantly to the absorption or emission on typical lines of sight. But if any such object could be identified it would be very interesting to study it in detail, and the topics aired in this paper are relevant to the physics of the solid matrix immediately below the surface of the body.

\subsection{Lifetime of the excited states}
In \S4 we evaluated the radiative width, $\Gamma$, of the excited states in each of the configurations under consideration, with the results presented in tables 1, 2 and 3. However, for \htwo\ molecules in a condensed environment it is also possible for the rovibrational energy of the \htwo\ molecule to flow into the many vibrational modes of the larger complex, rather than into radiation. That mode of decay has the potential to be rapid because the \htwo\ vibration frequency is very high ($\sim10^{14}\,{\rm Hz}$), so even a tiny coupling between the rovibrational modes of the \htwo\ molecule and the vibrational modes of the broader complex would be enough to make internal conversion the dominant de-excitation mode. For example: a 1\%\ probability of decay per oscillation would correspond to a state lifetime $\sim 1\;{\rm ps}$, whereas the largest value of $\Gamma$ to be found anywhere in tables 1-3 corresponds to a radiative lifetime about seven orders of magnitude longer.

When de-excitation occurs via internal conversion the energy could flow down a variety of channels. For example: if the \htwo\ molecule is adjacent to an \hsix\ molecular ion the mid-infrared vibrational modes of the latter could be excited. If so, some of the energy absorbed from starlight will reappear as mid-IR line emission from the molecular ion. That is a different pathway to energising the mid-IR emission of \hsix\ than the electronic excitation envisaged by \citet{lin2011}. Similarly, if the electrified \htwo\ molecule is part of a dust crystal then, regardless of the electric field configuration, energy can flow into lattice vibrations, thus heating the crystal --- a mechanism for powering the far-IR emission of solid \htwo\ dust from starlight.

\subsection{Diverse electrical environments}\label{sec:diverse}
As noted in \S5.1, electrified \htwo\ in the ISM could reside in a variety of microscopic electrical environments, and that has implications for how we might observe it. At the very least we are \emph{not} looking for a spectrum such as the one shown in figure \ref{fig:hminusspectrum}, which corresponds to a unique electrical environment, but a sum of several such spectra with weights appropriate to their relative incidence. That type of composite spectrum might arise from nanoclusters of \htwo\ when the thermodynamics strongly prefer certain configurations. In particular if solvation is only marginally possible then the fraction of ions would be a rapidly decreasing function of the number of \htwo\ ligands; in that circumstance the composite spectrum might be adequately represented using just the absorption spectra for the one- and two-ligand cases. 

However, as nanoclusters may incorporate a large number of ligand \htwo\ molecules, and each different coordination number implies a slightly different electrical environment (even for the first solvation shell, say), it seems more appropriate to consider the larger nanoclusters as providing a continuum rather than a small set of discrete possibilities. And of course, as commented earlier, charged dust crystals of \htwo\ manifest a continuum of electrical environments. In turn that means that each of the transitions of electrified \htwo\ in the ISM should present itself as a broad absorption band rather than a line. Quite how broad can be appreciated by referring to figure \ref{fig:electrifiedspectra} and noting the change in Stark-shift for each transition between the weakest field that might be encountered ($R,R_{eq}\rightarrow\infty$) and the strongest --- e.g. $R\simeq 2.8\;{\rm \AA}$ for the model of solvated \hminus\ discussed in \S4.3. Depending on the field configuration and the transition under consideration that width would be typically in the range $10^2 - 10^3\;{\rm cm^{-1}}$. 

Such numerous, broad absorption bands would overlap and would in effect contribute to the continuum extinction on any line-of-sight. Unfortunately that signature would be very difficult to distinguish from other continuum extinction contributions --- e.g. the scattering of light from dust particles. A more promising avenue for identifying electrified \htwo\ is the intraband structure. In other words: each absorption band has a spectral profile that may include some narrow structure at certain wavelengths. Importantly, there can be circumstances in which the transition wavenumber is stationary with respect to changes in the electrical environment, leading to an absorption ``caustic'' -- having the appearance of a spectral line -- within the absorption band. In fact numerous instances of turning points ${{\rm d} \lambda}/{{\rm d} R}=0$ can be found in figure \ref{fig:electrifiedspectra}, and a single example is clearly demonstrated in figure \ref{fig:pollhuntfundamental} (red line) at $R^{-3}\simeq 0.026\;{\rm \AA}^{-3}$ ($R\simeq 3.4\;{\rm \AA}$).

Not all of the transitions shown in figure \ref{fig:electrifiedspectra} exhibit turning points. Unfortunately there appears to be no simple way of anticipating the wavelengths or the specific circumstances under which stationary points will occur: one simply has to calculate. However, it is clear that turning points will not occur in weak fields, because in that case the eigenstates are close to those of the unperturbed molecule and the Stark shifts of all the levels scale in the same way with the strength of the perturbation. For example: at large separations, $R$, between molecule and ion the transition shown in figure \ref{fig:pollhuntfundamental} demonstrates a simple power-law behaviour of the Stark shift ($\propto R^{-3}$) because the quadrupole interaction dominates.

We emphasise that the electrostatic configurations modelled quantitatively in \S\S3,4 of the present paper are, however, idealised cases. In reality the field would include contributions from all the neighbouring molecules in the condensed complex (\S5.1), and any stationary points would exist within a much larger configuration space. For example: if the calculation is restricted to fourth-rank response tensors, as in the present work, we need 25 parameters to specify the electrical configuration --- i.e. the coefficients of a multipole expansion of the potential up to the hexadecapole. In practice the monopole is irrelevant (because \htwo\ has zero net charge), as are two of the dipole coefficients (choice of field direction is arbitrary), so only 22 parameters are required. Even so, that is a large space and a key issue affecting the prominence of any absorption lines is how much of that parameter space is sampled by the condensed \htwo\ molecules in the ISM.

On that topic it is worth noting that there is at least one situation where the occupied region of the configuration space is effectively one-dimensional, namely the circumstance of a uniform applied field. In that case all of the 22 field coefficients will be non-zero (in contrast to our description in \S\S3,4), as a result of the permanent and induced multipoles on each molecule in the crystal, but each coefficient depends only on the applied field strength. In this circumstance, with only a single control parameter, any stationary points should yield readily visible absorption lines.

It is important to recognise that the sort of spectrum we have just described -- in which lines form only as a result of stationary points -- may look very different to figure \ref{fig:hminusspectrum}, say. For example: the fundamental vibrational line is easily the strongest feature in that spectrum; but if there is no stationary point for that transition, across a broad range of sampled electrical environments, then there will be no absorption line. Instead the transition will simply appear as a broad band of absorption, and despite its strength it might be difficult to detect. Clearly, accurate predictions for the properties of stationary points will be required in order to identify electrified \htwo\ in the ISM.

\subsection{Emission from condensates}
In \S4.2.4 we pointed out the existence of a pure orientational transition in \emph{ortho-}\htwo.  Because the upper level of this transition is close to the ground state, at least in the case of weak fields, it can be substantially populated by thermal excitation even at the very low temperatures relevant to condensed \htwo. Consequently this transition has the potential to be an important cooling line for charged \htwo\ condensates. That is particularly true for nanoclusters, which have a limited number of modes of excitation; we expect this line to be less important as a coolant in the case of dust particles of condensed \htwo, because thermal continuum emission will be present in that case.

The wavelength of this orientational line exhibits a strong dependence on the strength of the electric field, as can be seen in figure \ref{fig:electrifiedspectra}. We have already noted (\S5.3) the likely diversity of electrical environments in the ISM, and because of that diversity the orientational transition can be expected to appear as a low-frequency bump on top of the thermal continuum of any charged \htwo\ dust particles. As such it may be of interest in connection with the so-called ``anomalous microwave emission'' (AME) that is observed from dusty gas clouds in the Galaxy \citep[e.g.][]{dickinson2018}.

Providing the condensate temperature is high enough that the Boltzman factor for the upper level is of order unity, it is straightforward to estimate the surface brightness of this orientational transition in terms of $N_{ortho}$, the column density of suitably electrified \emph{ortho-}\htwo:
\be
I_\nu\sim {1\over 3} \, \hbar A_{21} \,N_{ortho},
\ee
where we have assumed that the ground state is the $m=0$ state. Table 1 shows that in a uniform field with $R_{eq}=8\,{\rm \AA}$ the pure orientational transition has a frequency $\simeq 50\;{\rm GHz}$ and it proceeds at a rate $A_{21}\sim 3.3\times10^{-11}\,{\rm Hz}$; we adopt this transition rate as representative, for electrical environments that yield a line frequency of several tens of ${\rm GHz}$. Now the observed brightness temperature of the AME is approximately $10^{-5}\,{\rm K}$ at a frequency of $25\;{\rm GHz}$ and a gaseous hydrogen column of $10^{20}\,{\rm cm^{-2}}$ \citep[see figure 15 of][]{planck2014}, and this surface brightness would require $N_{ortho}\sim 2\times 10^{17}\,{\rm cm^{-2}}$. The fraction of \htwo\ in \emph{ortho} form is expected to be small, but its value is likely set by non-thermal processes and is difficult to predict. At a notional 1\% \emph{ortho} fraction, the total column of electrified \htwo\ needed to generate the observed level of AME is about 20\% of the gaseous hydrogen column.

The thermal continuum emission itself will also be different for charged versus uncharged \htwo\ dust particles. The imaginary part of the dielectric constant for pure solid \emph{para-}\htwo\ is very small at low frequencies \citep[][]{kettwich2015}, because the rotational transitions of \htwo\ are very weak. As we have seen, those transitions can be many orders of magnitude stronger when the molecule is electrified, leading to much more efficient thermal continuum emission. In turn this makes solid \htwo\ more plausible as an interstellar dust candidate, because radiative thermal equilibrium can be achieved at much lower temperature.

\subsection{The Diffuse Interstellar Bands}
Many of the spectra we exhibited in \S4 are entirely unlike the appearance of the molecule in zero field. In particular the dense forest of lines in the near-IR and optical bands that can be seen in figure \ref{fig:hminusspectrum} would be difficult to assign to \htwo\ if one did not have the appropriate theoretical template to hand. As the spectra in \S4 are the first attempt at constructing such theoretical templates it is likely that any optical absorption lines of electrified \htwo, had they been observed in the ISM, would have been incorrectly assigned or simply not assigned. As is well known, there is indeed a large number of interstellar optical/near-IR absorption lines that have not yet been assigned, known collectively as the Diffuse Interstellar Bands (DIBs) \citep[e.g.][]{herbig1995,sarre2006,camicox2014}; it is therefore of interest to consider whether the DIBs might be a manifestation of electrifed \htwo.

\subsubsection{Diffuse character of the absorption lines}
The large linewidths of the DIBs are a defining characteristic of the phenomenon. In the case of electrified \htwo\ short lifetimes for the excited states arise naturally because the molecules are part of a condensed system --- as discussed in \S5.2. In that discussion we considered a notional lifetime of the excited state of $\sim1\;{\rm ps}$, and in that case the implied linewidth is $\sim10\;{\rm \AA}$ for a transition in the violet part of the spectrum. That is comparable to the width of $12.3\;{\rm \AA}$ reported by \citet{jenniskens1994} for the strong DIB at $\lambda_{air}=4{,}428\;{\rm \AA}$. 

\subsubsection{Number and distribution of transitions}
The survey spectra reported in \S4 exhibit hundreds of strong transitions for each field configuration and each nuclear spin state. Thus, even though we are dealing with the simplest of molecules, the level of spectral complexity exhibited by electrified \htwo\ appears sufficient to account for the hundreds of DIBs that are now known \citep[e.g.][]{jenniskens1994,hobbs2008,hobbs2009}. Only a handful of the known DIBs lie in the infrared at wavelengths $\lambda\ga 1\;{\rm \mu m}$ \citep[][]{joblin1990,geballe2011}, and electrified \htwo\ also shows relatively few transitions in that region (figure \ref{fig:electrifiedspectra}). 

At the other end of the spectrum, no DIBs are known shortward of about $0.4\;{\rm \mu m}$ \citep[e.g.][]{jenniskens1994,herbig1995}. The lack of DIBs in the far-UV \citep[][]{snow1977,seabsnow1985} causes no difficulty because such photons lie in the \htwo\ rovibrational continuum. However, in the near-UV the situation is different: all of our models are rich with transitions down to at least the atmospheric cutoff at around $0.3\;{\rm \mu m}$ and that is a clear discrepancy between model and data. It may be possible to reconcile this difference when allowance is made for the lifetime of the excited states, as discussed in \S5.2: higher levels of excitation correspond to larger internuclear separations of the \htwo, leading to stronger interactions with near-neighbours and thus a faster redistribution of energy within the vibrational modes of the condensed complex. In other words: it is possible that near-UV DIBs do exist but have not been detected because they are much broader than the optical and near-IR DIBs. 

\subsubsection{Line assignments}
In the context of interstellar absorptions we have already noted that electrified \htwo\ is likely to give rise to very broad bands, with ``lines'' appearing only where the transition wavelength is stationary with respect to the field parameters. In principle the stationary points can be determined and the properties of the resulting lines can be compared to those of the DIBs. However, as discussed in \S5.1 (see also \S6.1), the models presented in this paper are rather crude representations of the real systems and we are not in a position to undertake meaningful calculations. 
 
\subsubsection{Line strengths}
It is useful to establish an estimate of the quantity of material that would be required to explain the DIBs. If the column-density of electrified \htwo\ molecules is $N_{mol}$, foreground to a particular source, then the resulting dimensionless equivalent width ($W\simeq W_\lambda/\lambda_0$) of a line with wavelength $\lambda_0$ is
\be
W\sim N_{mol}\,{{\lambda_0^3}\over{8\pi c}}\, A_{21},
\ee
where we have neglected the (of order unity) ratio of degeneracies of the upper and lower levels. At a reddening of $E(B-V)= 1$, a moderately strong DIB with $W_\lambda\sim 50\,{\rm m\AA}$ and  $\lambda_0\sim5{,}000\,{\rm \AA}$ would require a column $\sim 6\times 10^{19}/A_{21}\;{\rm cm^{-2}}$. That figure is for the molecules that are responsible for the absorption line. But if the line is showing us only those absorptions whose wavelengths are close to a stationary point then the total column density will be much larger; for the purpose of obtaining numerical estimates we assume that the total column is ten times larger.

Although electrification leads to \htwo\ rovibrational lines that are orders of magnitude stronger than in the field-free case, they are nevertheless still quite weak. Taking $A_{21}\sim10^{-2}\,{\rm Hz}$ as an estimate of the transition rate for one of the stronger transitions of electrified \htwo\ -- see figure 9 -- we arrive at $N_{mol}\sim 6\times 10^{22}\;{\rm cm^{-2}}$ for $E(B-V)= 1$ ($A_V\simeq3$). We return to this result in \S\S6.3,6.4.

\subsubsection{Variations in DIB line ratios}
It is now firmly established \citep[][]{cami1997,mccall2010,friedman2011} that the pairwise correlation coefficients between DIBs are broadly distributed and sometimes quite low (e.g. $\la 0.5$); they are rarely stronger than 0.95. By contrast one would normally expect near-perfect correlations amongst all lines arising from the ground state of a given carrier, so the poor correlations that are observed have to date been interpreted as a requirement for multiple species of carrier. In that case a large number of different carriers, comparable to the number of observed DIBs, is implied. In the case of electrified \htwo, however, we have argued that observed absorption \emph{lines} arise from particular electrical environments where the wavelength is stationary with respect to small changes in those environments. Each such line arises from a different stationary point, corresponding to a different environment, and the observed line strengths therefore reflect the different populations of the same molecule in the vicinity of the various different points. Those populations are not expected to be tightly correlated, so in the case of electrified \htwo\ \emph{all} of the DIBs could in principle arise from just the one carrier.

The degree to which populations in the various different electrical environments might be correlated will depend strongly on the true dimensionality of the electrical configuration space. As noted in \S\ref{sec:diverse}, 25 parameters are required to uniquely specify the electrical environment up to the hexadecapole of the potential, but for many physical models there are strong correlations between those parameters and consequently the true dimensionality of the electrical parameter space can be much smaller. For example: in the case of a uniform field applied to a crystal of \htwo\ all of the potential multipoles are fixed by the applied field strength, so the parameter space is only one dimensional. In that case the strength of the correlation between a pair of lines ought to reflect mainly the difference in field strength of the two stationary points, because the grains ought to be smoothly distributed in field strength. In order to predict the DIB-DIB correlation matrix, however, it will be necessary to identify both the general character of the electrical environments -- e.g. a crystal with a uniform applied field, or an \htwo\ nanocluster centred on a single ion -- and the astrophysics that governs the statistics of those environments.

\section{Discussion}
\subsection{Limitations of the model calculations}
Our calculations provide a more accurate characterisation of electrified \htwo\ than has been available to date \citep{pollhunt1985}; but our model is nevertheless far from ideal in some respects. First, although the inclusion of all the fourth-rank response tensors has only recently become possible -- and thus represents the state-of-the-art for calculations of this type -- there is no reason to think that fourth-rank suffices; rather the opposite in fact. A comparison between second-rank and second- plus fourth-rank response is shown in figure~\ref{fig:pollhuntfundamental} (\S3.3) for the particular case of the wavenumber of the $\ket{\hat{0},\hat{0},0}\leftrightarrow\ket{\hat{1},\hat{2},1}$ transition for \htwo\ near a positive elementary charge. In that case we can see that the fourth-rank contribution is small compared to the second-rank contribution only for separations $R\gg2.7{\rm\AA}$. However, in the case of the \hsix\ ion, for example, the calculations of \citet{kurosaki1998} indicate that the first \htwo\ solvation shell has a radius of approximately $2.7{\rm\AA}$. Furthermore, as noted in \S3.2, the fourth-rank response tensors typically become more important relative to the second-rank contributions as the level of excitation increases. Consequently the sixth- and higher rank response tensors would likely be even more important if we are seeking an accurate description of optical absorption lines, rather than the fundamental vibrational line shown in figure \ref{fig:pollhuntfundamental}.

The situation we have just described is likely to be common to the first solvation shells of other small ionic species, regardless of the sign of the charge. For example in \S4.3 we presented a model for \htwo\ in the vicinity of the \hminus\ ion, designed to match the measured wavenumber of the fundamental vibrational line, for which the preferred separation was determined to be $R\simeq2.8{\rm\AA}$. Thus we expect the fourth- and higher-rank tensor contributions to be important in this context also.

Unfortunately the sixth-rank response tensors are even more numerous than the fourth-rank --- a total of eleven additional tensors, including the permanent 64-pole moment, the octupole polarizability, and the fourth dipole hyperpolarizability, etc. They would also be more challenging both to evaluate \emph{ab initio} and to work with in modelling the eigensystem of electrified \htwo. As these difficulties would be further magnified for the higher rank response tensors (i.e. rank 8 and above), further expansion of the electrical response model does not seem to be a promising approach to achieving higher accuracy in the strong electric fields under consideration here.

A second concern is that the bound rovibrational states of the unperturbed system may not suffice to describe the states of the electrified molecule, i.e. our basis is incomplete. This concern is justified, as noted in \S4.2.3: the energy eigenvalues become systematically more negative with increasing field strength, and in strong fields a gap opens up between the least-bound state and the continuum. This suggests that states in the unbound continuum of the field-free molecule may become bound when a field is applied.

Both of the aforementioned problems have their origin in the fact that the electric fields under consideration include quite large values, for which the electrical energy is not really a small perturbation. It therefore seems likely that a different calculation framework would be more appropriate for further investigation of the properties of electrified \htwo, at least at the upper end of the range of field strengths that we have considered. That idea is strongly reinforced by the fact that our description is purely electrostatic and completely neglects all other interactions between the \htwo\ molecule and an adjacent ion. By contrast all of the methods of theoretical quantum chemistry are designed to correctly describe such interactions and the strong electric fields that are encountered. What is needed now is quantum chemistry rather than quantum physics. 

Notwithstanding the limitations of the model we have used, a great merit is that it furnishes us with a (nearly) complete set of rovibrational transitions for electrified \htwo. Quantum chemical calculations, on the other hand, have to date been more oriented towards understanding aspects such as the binding energy, the geometry, and the fundamental vibrational and rotational transitions --- high levels of rovibrational excitation are not routinely characterized.

\subsection{Absorptions into the continuum}
All of the transitions reported in \S4 involve upper levels that are bound states of the electrified \htwo\ molecule, but presumably there are also allowed transitions that lead to states in the rovibrational continuum. We have not attempted to characterize those transitions so we cannot say much about their properties, but it is interesting to consider where the threshold of that process might lie.

To be specific we take the case of the hydrogen anion with a single, complete solvation shell, i.e. ${\rm H}^-({\rm H}_2)_{12}$. Absorption of a UV photon can directly dissociate one of the ligand \htwo\ molecules, via the reaction
\be
{\rm H}^-({\rm H}_2)_{12} \;\;\; + \;\;\; \gamma\;\;\;  \rightarrow\;\;\;  {\rm H}_2^*({\rm H}_2)_{11} \;\;
+ \;\; {\rm H}^-.\label{eq:dissociation}
\ee
In this process the polarisation of the dissociating molecule increases to a point where the dissociation products are ${\rm H}^+$ and \hminus; the latter escaping from the complex and the former bonding with the central \hminus\ to yield ${\rm H}_2^*$, i.e. a highly vibrationally excited \htwo\ molecule. The ${\rm H}_2^*$  will be close to the dissociation limit, because the constituent atoms have come together from an initial, widely separated state, so its energy relative to the ground state will be approximately $4.5\;{\rm eV}$. The \hminus\ that escapes the complex has some kinetic energy; but that ought to be small near the threshold of the process.

The \htwo\ molecules that are clustered around the ${\rm H}_2^*$ will also disperse to infinity, because once the charge has been removed from the centre they are all unbound. The interaction energy can be estimated from the pairwise interaction potential evaluated at their initial separation --- i.e. at the radius of the first solvation shell around \hminus, which is approximately $2.8\;{\rm \AA}$ (\S4.3). From the isolated pair interaction potential of figure 4 in \citet{silvera1978} we estimate the total potential energy of the eleven remaining molecules as roughly $+0.2\;{\rm eV}$. The total energy of the reaction products is thus approximately $+4.7\;{\rm eV}$.

By contrast the total energy of the ${\rm H}^-({\rm H}_2)_{12}$ cluster is negative: the cluster is bound. The results of \citet{mohammadi2020} indicate a binding energy per ligand of approximately $1.4\;{\rm kcal\,mol^{-1}}$ for ${\rm H}^-({\rm H}_2)_5$, which is the largest cluster they studied. Adopting that figure for all twelve of the ligands in the first solvation shell then implies a total binding energy of approximately $0.7\;{\rm eV}$. We therefore require a photon whose energy is at least $5.4\;{\rm eV}$ in order to initiate the reaction (\ref{eq:dissociation}). That threshold energy corresponds to $\lambda\simeq2{,}300\;{\rm \AA}$: close to the central wavelength of the UV bump ($2{,}175\;{\rm \AA}$) that is commonly seen in the interstellar extinction curve \citep[e.g.][]{draine2003}, and the reaction (\ref{eq:dissociation}) may be of interest as a possible contribution to that feature.

\subsection{Electrified \htwo\ as a DIB carrier?}
In \S5.4 we proposed electrified \htwo\ as a DIB carrier, on the basis of several features of the absorption system that are in accord with the properties of the DIBs, but some potential difficulties for that proposal come to mind, as we now describe. 

\bigskip
\subsubsection{Multiple carriers versus just one}
A great variety of carriers have been proposed for the DIBs, as discussed in, e.g.: \citet{herbig1995,sarre2006,camicox2014}. There has even been a prior suggestion that \htwo\ might be responsible for many of the lines \citep{sorokin1995,sorokin1996}, although \citet{snow1995} has raised objections to some aspects of that proposal. Almost all of the suggested carriers have subsequently been eliminated by dint of a spectral mismatch between laboratory and astronomical data. But there is one candidate molecule, $C_{60}^+$, the Buckminsterfullerene cation -- suggested by \citet{foing1994} --  that remains viable. Recent data from both laboratory and telescope have strengthened the case in favour of $C_{60}^+$ being a DIB carrier, with multiple transitions exhibiting a close wavelength correspondence to near-IR DIBs \citep[][]{campbell2015, cordiner2019}. How, then, should we view those assignments in the context of the broader DIB interpretation being advanced in this paper?

At present there is no conflict between electrified \htwo\ and the $C_{60}^+$ identifications, because we do not have accurate predictions of line locations and strengths for the former. It is possible that electrified \htwo\ will prove unable to reproduce the locations and strengths of the DIBs that have been assigned to $C_{60}^+$, making the latter species complementary to the former. In that case the result would be a more complex picture than if all the DIBs could be explained by electrified \htwo, but there is no fundamental difficulty with such a composite model. Furthermore the number of DIBs that can potentially be attributed to $C_{60}^+$ is less than 1\%\ of the total number known, so the motivation for considering electrified \htwo\ is hardly diminished at all. Indeed that would remain true while the origin of the bulk of the DIBs remains uncertain, even if additional carriers were identified in future. In that case electrified \htwo\ would be simply one carrier amongst many, albeit one with the potential to account for a large number of lines.

It also remains a possibility that $C_{60}^+$ will turn out not to be responsible for the DIBs that have been attributed to it. There are reasons to be cautious about the assignment. First, the astronomical data exhibit substantial structure in the ``continuum'' in the vicinity of the DIBs in question, making it difficult to measure line centroids and equivalent widths accurately. Secondly, because DIBs are very numerous and their profiles are broad, individual wavelength coincidences have a non-negligible probability of arising purely by chance. A simultaneous match for several wavelengths, as demonstrated for $C_{60}^+$, is highly improbable when considered in isolation; but it becomes much more likely when viewed as the outcome of a process in which a large number of candidate carrier species are considered, and unsuccessful candidates are discarded. These comments are not intended to disparage the $C_{60}^+$ assignments, but to encourage readers to give serious consideration to the possibility that all of the DIBs might be attributable to a single carrier.

\subsubsection{Interpretation of diffuse band profiles}
Although some DIBs are consistent with Lorentzian profiles, others are known to show substructure. For example the $6{,}614\,{\rm \AA}$ DIB shows multiple peaks with small separations; these have been interpreted as rotational substructure arising from a large molecule with a  rotational constant in the range $0.003\la B({\rm cm^{-1}})\la 0.02$ \citep[much smaller than that of \htwo;][]{sarre1995,kerr1996}. The question then arises as to how the $\sim 1\;{\rm cm^{-1}}$ splitting that is observed for the $6{,}614\,{\rm \AA}$ DIB, for example, might be interpreted under the hypothesis that electrified \htwo\ is the carrier? Certainly the scale of the substructure is not comparable to either the rotational or vibrational line separations of \htwo\ -- which are hundreds or thousands of ${\rm cm^{-1}}$, respectively. One possibility is that electrified \emph{ortho-}\htwo\ lines might exhibit small splittings due to hyperfine interactions between the nuclear spin $I=1$ and the nuclear angular momentum, $j$; such interactions are not included in our Hamiltonian (\ref{eq:electricalenergy}). Thus, although we have not demonstrated transition multiplicity akin to that seen in some of the DIB data, it should not be assumed that electrified \htwo\ cannot exhibit such splittings.

\subsubsection{DIBs correlate strongly with {\rm H{\sc i}}}
The fact that DIB equivalent widths are known to correlate strongly with the atomic hydrogen column-density, and only weakly with the (gaseous) molecular hydrogen column-density \citep[e.g.][]{herbig1995} is an obvious issue that needs to be faced up to. There are, however, some good reasons to think that this is not a fundamental incompatibility. First, in the ISM electrified \htwo\ can only occur in condensates, not in the gas phase, and it is not at all clear what relationship should be expected between the column-densities of \htwo\ in the two forms. As they are different thermodynamic phases of the same substance one might even expect there to be an anticorrelation, with the dominant form being determined by the local thermodynamic conditions. Secondly, as emphasized in the Introduction, any gaseous \htwo\ that is cold and dense enough to permit precipitation of the pure solid is likely to have escaped detection to date \citep[e.g.][]{pfenniger1994a,gerhard1996,walkerwardle2019}. In other words: at present we might not be measuring the largest contribution to the gaseous molecular hydrogen column. Thirdly, we recall the suggestion by \citet{allen2004} that diffuse atomic hydrogen may arise as a result of UV-photodissociation of uncatalogued \htwo. That idea seems particularly attractive in view of the large column of electrified \htwo\ that we estimated would be required to explain the measured DIB strengths (\S5.5.4, but see also \S6.4). 

In addition to the strong correlation between DIB equivalent widths and {\rm H{\sc i}}, weaker correlations with \htwo\ have been demonstrated by \citet{lan2015}, with sufficient precision to reveal different dependencies for different lines. It was shown by \citet{lan2015} that modelling DIB strengths as power-laws in both N({\rm H{\sc i}}) and N(\htwo) yields a compact description that captures several aspects of DIB phenomenology --- phenomenology that had previously been described in different terms. To connect that description to the present paper we suggest that the two gas column-densities are not themselves the fundamental variables that determine DIB strength, but that those column-densities provide gauges of the interstellar conditions that do determine DIB strength. In particular we note that the UV radiation field is expected to play an important r\^ole in grain charging, as well as in converting \htwo\ to  {\rm H{\sc i}}, so the key ingredients could be the UV field and a supply of condensed hydrogen.

\subsubsection{A solid-state carrier?}
The possibility that the DIBs might arise in dust grains is not a new idea. Quite the opposite in fact: the correlation between dust-reddening and DIB equivalent widths made solid-state carriers an early focus of attention in the history of DIB research \citep[e.g][]{herbig1995}. But the prevailing view at present is that the carriers are more likely to be molecules in the gas phase. \citet{snow2004} summarized the main evidence against a solid-state carrier of the DIBs in four key points: (i) the DIBs show little or no profile or wavelength variations, even amongst lines-of-sight showing a variety of dust extinction characteristics; (ii) the DIBs show none of the ``emission wings'' that were anticipated for absorption lines arising from dust particles; (iii) there is little or no variation in polarisation across individual DIBs; and, (iv) some of the DIBs are much narrower than typical solid-state transitions. 

The last of those points is clearly not a compelling argument because there is no requirement for the DIBs to be ``typical solid-state transitions.'' In fact solid \emph{para-}\htwo\ is far from being a typical solid; in its pure form it is used for matrix isolation spectroscopy of small molecules precisely because the linewidths can actually be even narrower than in the gas phase \citep[e.g.][]{fajardo2009}.

The difficulties posed by points (i) and (ii) only apply to a subset of interstellar dust particles --- depending on the refractive index and particle size/shape. In our case the relevant refractive index is unknown, but a useful fiducial refractive index is that of pure, solid \emph{para-}\htwo, which in the optical has the approximate value $1.13$ with a very tiny imaginary component \citep[][]{kettwich2015}. For small particles with refractive index close to unity Rayleigh-Gans theory applies, and spheres with radius small compared to the wavelength exhibit absorption line profiles that are indistinguishable from molecules in gas phase \citep[section 11.41 of][]{vandehulst1981}. Not just spheres, in fact: for small particles with refractive index close to unity the influence of shape is simply that of a ``form factor'' which affects the overall normalisation, but not the spectral profile of an absorption line \citep[chapter 6 of][]{bohren2004}. For this entire class of hydrogen dust particles, then, no variation in the wavelength or profile of absorption bands can arise from the particulate nature of the carrier. Similarly, for that class of particles, there is no apparent emission on the wings of the line. Small particles also dodge the polarisation bullet (iii): \citet{whittet2004} states that the observed polarisation of starlight arises almost entirely in the large grain population.

The arguments mustered above demonstrate that there is at least one possible manifestation of \htwo\ condensates that satisfies the known constraints on solid-state DIB carriers --- namely sub-wavelength particles. It is possible that there might be other manifestations of solid \htwo\ that could meet the known constraints, but it is beyond the scope of this work to explore that possibility. 

\subsubsection{Relationship of the DIBs to other observables}
A tighter grain size limit can be determined by considering the implied extinction due to scattering. In \S5.5.4 we estimated that a column $N_{mol}\sim6\times 10^{22}\;{\rm cm^{-2}}$ of electrified \htwo\  would be required to generate the DIB strengths seen at a visual extinction of about 3~magnitudes. For spherical hydrogen grains of radius $a$ this molecular column translates into a grain column of $N_{grain}({\rm cm^{-2}})=(a/0.82\;{\rm cm})^{-3}$. And if the real part of the dielectric constant is similar to that of pure, solid  \emph{para-}\htwo\ \citep[$\epsilon\simeq 1.27$,][]{kettwich2015} then each grain has a scattering efficiency (i.e. cross-section in units of $\pi a^2$) of $Q_{sca}\simeq 30(a/\lambda)^4$ \citep[eq. 5.8 of][]{bohren2004}. It follows that $A_V<3$ requires $a\la 100\,{\rm \AA}$. We note that this limit does not automatically prefer isolated nanoclusters over dust crystals, because the smallest individual nanoclusters have radii $\simeq 3\,{\rm \AA}$. 

In view of the low levels of optical absorption in pure solid \emph{para-}\htwo, the transitions giving rise to the DIBs should dominate the optical absorption in charged grains. Consequently the measured properties of the DIBs themselves can be used to estimate the resulting absorption. We use a Lorentz Oscillator model to approximate\footnote{As noted in \S6.3.2, not all of the DIB profiles are Lorentzian.} the imaginary part of the dielectric constant of the ISM in terms of the measured properties of the known DIBs: the central frequency, $\nu_j$; the corresponding FWHM, $\gamma_j$; and, the dimensionless equivalent width, $W_j$. That model gives a total absorption optical depth of
\be
\tau(\nu) = {6\over\pi}\; {\rm Im} \; \sum_j {{W_j\, \nu_j\,\nu}\over{\nu_j^2 - \nu^2 - i\,\nu\gamma_j}},
\ee
with no adjustable parameters.  Taking the properties of the 378 DIBs reported by \citet{hobbs2008}, which correspond to a line-of-sight with $E(B-V)\simeq1$ ($A_V\simeq 3$), we find that the absorption in $V$-band has a floor at $\tau\sim10^{-4}$. That is too small to account for a continuum extinction of $A_V\simeq 3$. However, in our preferred interpretation each of the DIBs is only a fraction of a stronger and \emph{much} broader absorption band of electrified \htwo. To obtain a rough, quantitative estimate we assume that (i) each observed DIB manifests only $\sim$10\% of the total strength of each absorption band, and (ii) the full width of each band is $\gamma_j\sim\nu_j/20$. With these assumptions the $V$-band optical depth is a few percent: still too small to explain the continuum extinction that is observed, but not completely negligible.

The foregoing estimates suggest that there may be scope for accommodating electrified hydrogen as an addition to conventional dust models based on silicates and carbonaceous grains, with the former accounting for the DIBs and the latter accounting for the observed continuum extinction. But a simpler model might be possible, in which charged hydrogen grains also account for the bulk of the continuum extinction. That simplicity is appealing, but a viable dust model based exclusively on \htwo\ condensates has not yet been demonstrated, and the requirements for a successful model go well beyond just extinction --- see, e.g., \citet{draine2003}.

We have already noted (\S5.4) a possible connection between the pure orientational transition of electrified \emph{ortho}-\htwo\ and the anomalous microwave emission that is observed from dusty regions of the Galaxy. But it is known that AME is tightly correlated with the far-IR thermal emission from dust \citep[e.g.][]{hensley2016}, so ideally one would like to have a model in which both of these radiations arise from the same particles. Thus a better understanding of charged \htwo\ grains would be of interest not only in connection with the DIBs, but also the continuum extinction, AME and the far-IR emission.

\subsection{Total mass of electrified \htwo}
In \S5.5.4 we estimated that $N_{mol}\sim6\times10^{22}\,{\rm cm^{-2}}$ in charged hydrogen grains would be required to explain the observed DIB strengths at a reddening of $E(B-V)=1$. By contrast the expected column of diffuse gas at this reddening is much smaller: $N_{\rm H}\simeq6\times10^{21}\,{\rm cm^{-2}}$ \citep[e.g.][]{draine2003}. This remarkable situation, in which the dust outweighs the gas by a factor $\sim20$, is opposite to the hierarchy that applies in conventional (silicate + carbonaceous) dust models, where the grains amount to only $\sim 1$\% of the gas mass. As can be seen from the way $N_{mol}$ is calculated in \S5.5.4 the large total grain mass that we infer follows directly from the fact that, despite electrification, the rovibrational transitions are not very strong.

Although such a large mass in dust is initially surprising it is not entirely unwelcome. Galaxies contain a great deal more mass than can be seen in stars and diffuse gas \citep[e.g., the many contributions to][]{ryder2004}, so there is certainly scope in the dynamics for a substantial mass in \htwo\ grains. Furthermore the fact that DIB strengths are well correlated with the H{\sc i} column (\S6.3.3) means that the contribution of the DIB carriers to galaxy rotation curves is simply a scaled version of the contribution of the H{\sc i} itself, and it is known that much of the observed structure of rotation curves can be accommodated by simply scaling the contributions of the stars and the diffuse gas \citep[][and references therein]{swaters2012}.

Despite its dynamical appeal we can rule out the possibility that such a large mass resides in small particles of condensed \htwo. Material in the ISM is bombarded by cosmic rays, resulting in $\gamma$-ray emission via bremsstrahlung and pion production, and these diffuse, Galactic emissions have been studied by a succession of satellite instruments since the 1970's \citep[e.g.][]{kraushaar1972,hunter1997,ackermann2011}. The observed $\gamma$-ray intensities are similar to those calculated on the basis of the independently-measured cosmic-ray spectra and the spectral-line surveys of the molecular and atomic gas content of the Galaxy, and this similarity in absolute normalisation excludes the possibility that the interstellar dust in the diffuse ISM is twenty times more massive than the diffuse interstellar gas.

There are, however, some substantial systematic uncertainties involved in the calculation of diffuse $\gamma$-ray emissions, so this avenue of constraint does not oblige us to accept a dust mass that is as small as the $\sim1$\% (of the gas mass) that is usually assumed. Several factors entering into the calculation are uncertain by $\ga 10$\% \citep[][]{delahaye2011}, including: the relationship between the local interstellar cosmic-ray spectra and the spectra measured inside the heliosphere \citep{cummings2016}; the spatial variation of cosmic-ray spectra within the Galaxy \citep{strong2007}; the cross-sections for $\gamma$-ray production \citep{mori2009}; the optical depth of the 21cm line of atomic hydrogen \citep{ackermann2011}; and, the relationship of the total molecular column to the measured CO line intensity \citep{bolatto2013}. In addition to those modelling systematics, the instrumental calibration is only secure at the $10$\% level \citep{ackermann2012}. Consequently the $\gamma$-ray data can only provide limits on a massive dust component at the level of a few tens of percent of the gas mass.

A complementary approach to this issue is to recognise that the column in condensed \htwo\ will not be perfectly correlated with the gas column, and then to identify the $\gamma$-ray emission that is associated specifically with dust -- as measured by reddening, say -- rather than with the gas. That separation has already been made by \citet{grenier2005}, who inferred a dust-associated-mass component at a level $\sim 30$\% of the measured gas mass. The dust-associated-mass was interpreted by \citet{grenier2005} as dark gas, because the dust itself was assumed to be made up of metals and thus to be a negligible fraction of the mass. But the result could alternatively be interpreted as a large mass contribution in condensed \htwo.

In connection with these two alternative interpretations we note that gas clouds in which hydrogen can condense may have very high central column-densities \citep[e.g. $\sim 10^3\,{\rm g\,cm^{-2}}$ for the cloud shown in figure 3 of ][]{walkerwardle2019}. So high, in fact, that cosmic-rays cannot penetrate and the $\gamma$-ray emissivity is consequently very low --- see the calculations in \citet{ohishi2004}. Such clouds are not only ``CO-dark'' but also ``$\gamma$-ray dark'' and might not be revealed by analyses like that of \citet{grenier2005}.

The $\gamma$-ray constraints just described require the condensed \htwo\ mass to be $\la 60$ times smaller than we estimated in \S5.5.4 --- a discrepancy which effectively excludes that particular model. A viable model would require radiative transition rates that are one or two orders of magnitude higher than the $A_{21}\sim 10^{-2}\,{\rm Hz}$ assumed in \S5.5.4. That transition rate was based on our calculations for \htwo\ in the first solvation shell of an ion (figure 9), so there is little prospect of us reproducing the necessary line strengths within the confines of that particular electrical configuration. For the other electrical configuration that we have studied, i.e. the case of a uniform field, the maximum field strength that can arise from collisional charging of \htwo\ grains has previously been estimated as $F_{max}\simeq 1.6\times 10^{10}\,{\rm V\, m^{-1}}$ \citep{walker2013}, which corresponds to $R_{eq}\simeq 3.0\,{\rm \AA}$ --- hence a slightly weaker field than the case shown in figure 9, and thus similarly unpromising.

Although our models cannot yield the necessary DIB equivalent widths, while simultaneously remaining consistent with the diffuse $\gamma$-ray constraints, these are not grounds to reject the possibility of electrified \htwo\ as a DIB carrier. When compared to the physical circumstances that we are considering, the models we have constructed are highly simplified representations (\S6.1); more realistic descriptions are needed to address this issue. Although that statement is general, we note in particular that our descriptions of both electrical configurations neglect the fields arising from all of the neighbouring, polarized \htwo\ molecules. Those neighbours introduce substantial high-order multipoles into the electrical potential structure around the molecule of interest, with corresponding contributions to the induced dipole moment (cf. equation 8) and the associated transition rate.

\section{Summary and Conclusions}
We have investigated the rovibrational eigenstates and ground state rovibrational absorption spectra of \htwo\ in the X$\,^1\Sigma_g^+$ electronic ground state, when subjected to an external, static electric field. Our treatment uses recent \emph{ab initio} calculations of the static electrical response tensors of \htwo, complete up to fourth rank, and substantially improves upon the best previous characterisation in the literature. Results were obtained for a point-like charge $q=\pm e$ at 99 values of the separation $R$ in the range $8\ge R({\rm\AA}) \ge 2$, and for a uniform field at the same set of field strengths. For each configuration, all transitions stronger than $A_{21}=10^{-12}\;{\rm Hz}$, in either parallel or perpendicular polarisation, have been tabulated.

In all of the studied configurations the molecule exhibits absorption lines that are orders of magnitude stronger than any of its field-free rovibrational absorptions from ground-state, as a result of the electric dipole moment that is induced by the static field. Consequently the electrified molecule could feature prominently in astronomical spectra even though it may be only a tiny fraction of the total \htwo\ column on any line-of-sight.

The fact that the energy eigenstates of electrified \htwo\ are mixtures of both rotational and vibrational eigenstates leads to a large number of permitted transitions, so our calculated spectra are much richer than those of the field-free molecule. We find that mixing is much stronger amongst the vibrationally excited states than for the vibrational ground state, leading to a forest of optical absorption lines for molecules immersed in strong, static fields. These spectra bear little resemblance to the absorption spectra of the field-free molecule, so absent a suitable theoretical template it would be difficult to recognize the carrier as \htwo\ if any such spectrum appeared in astronomical data.

That point alone is enough to motivate consideration of electrified \htwo\ as a possible carrier of the DIBs, but we noted two further aspects of our model that reinforce the idea. First, if the electrified molecule were part of a condensed system -- as is expected in the interstellar context -- then its excited rovibrational states would couple to other vibrational modes of the system, so that rapid internal conversion of the excitation energy would be likely to occur, and this would give the optical absorption lines a diffuse character. Secondly, we find a rapid increase in the number of transitions as the wavelength decreases across the near-IR, with only a small number located longward of $\lambda=1\;{\rm\mu m}$ --- as is known to be the case for the DIBs.

The distribution of transitions with wavelength does, however, also manifest a clear mismatch between the model and the data, with the DIBs observed only down to $\lambda\sim0.4\;{\rm\mu m}$, whereas electrified \htwo\ exhibits a high density of transitions down to at least  $\lambda\sim0.3\;{\rm\mu m}$.

Considering the variety of microscopic environments that is likely to exist within interstellar condensates we argued that each transition of electrified \htwo\ ought to imprint absorption over a band that is very broad indeed --- much broader than any of the DIBs. That led us to propose a more specific interpretation: that the DIBs should be identified with points where the transition wavelength is locally stationary with respect to variations in the electrical environment. Because each such point arises from a different microscopic field configuration the resulting set of absorption peaks will not have tightly correlated equivalent widths, leading to the novel possibility that all of the DIBs might be explicable as absorptions from the ground state of this one carrier.

Despite a huge increase in transition strength relative to the field-free case, the lines of electrified \htwo\ are still quite weak, and the observed equivalent widths of the DIBs require a large column-density in charged \htwo\ grains, if those grains are indeed the carrier. We estimated their mass contribution to be larger than that of the diffuse gas, but such a large mass in \htwo\ grains is excluded by the available constraints on unmodelled, diffuse Galactic gamma-ray emission. Because the requisite number of carriers follows fairly directly from the transition strength, which in turn reflects the induced dipole moments, we speculated that this conflict might perhaps be resolved by more realistic descriptions of the electrical environment, accounting for the high-order multipolar fields due to neighbouring \htwo\ molecules --- fields which have been entirely neglected in this study. Even so the transition dipoles would need to be at least an order of magnitude larger than we have calculated in order to be able to explain the DIBs with trace quantities of condensate, and our suggestion of electrified \htwo\ as the DIB carrier does not sit easily with the current understanding of the ISM.

Amongst the new transitions that emerge from electrification we found an \emph{ortho-}\htwo\ line at very long wavelengths associated with a reorientation of the molecule. Because it is readily thermally excited, even at low temperatures, this transition is potentially important in astrophysical contexts as a cooling line for molecular hydrogen condensates. Observationally it ought to appear as a broad band, reflecting the range of electrical environments in which the molecules are situated, and we suggested a possible connection to the ``anomalous microwave emission'' detected from dusty interstellar clouds.

The fact that electrified \htwo\ has electric dipole transitions at low frequencies -- not just the reorientation transition for \emph{ortho-}\htwo, but also the pure rotational lines for both \emph{para-} and \emph{ortho-} sequences -- means that charged hydrogen dust will emit thermal continuum radiation far more efficiently than any uncharged grains of \htwo. Electrification may be important for both sides of the heating-cooling balance: we noted that thermal far-IR radiation from hydrogen grains can arise from the absorption of starlight in the bands of electrified \htwo, with the rovibrational energy of the latter undergoing internal conversion into phonons. A similar process provides a novel means of exciting suprathermal mid-IR line emission from \hsix, or \hdthree, when that ion is surrounded by \htwo\ ligands.

Although we have made a case that electrified \htwo\ has astrophysically interesting aspects, that case rests largely on the broad-brush characteristics of our calculated spectra. For example: although we have suggested that electrified \htwo\ may be a carrier -- perhaps even \emph{the} carrier -- of the DIBs, we are not in a position to support that proposal with specific predictions for the properties of the individual lines. Moreover, the transition probabilities we estimated are too small to allow trace quantites of electrified \htwo\ to account for the DIBs, making the proposal incompatible with the accepted picture of the ISM. However, the calculations presented herein are based on a highly simplified representation of the real physical system which does not accurately describe the strongest electric fields that are of practical interest, and does not account at all for either the fields due to neighbouring \htwo\ molecules or the mutual interaction of the electrons in the molecule with those in any adjacent moiety. Quantum chemical treatments are now needed, with a focus on understanding the full spectrum of rovibrational excitations of the \htwo\ moieties within larger, condensed systems. In parallel, new experimental investigations of the absorption spectra of electrified, condensed \htwo\ would be valuable --- both in the bulk, and in nanocluster form.

\begin{acknowledgements}
Thanks to Bob Brooks for his insights into the laboratory studies of irradiated hydrogens, and to Artem Tuntsov for many good discussions about the modelling and a careful reading of the manuscript. I'd also like to express my appreciation of some thoughtful refereeing. Access to the Bodleian Library was very helpful and the author is grateful to the Department of Astrophysics, Oxford, for their hospitality.
\end{acknowledgements}

\appendix
\begin{figure}
\plottwo{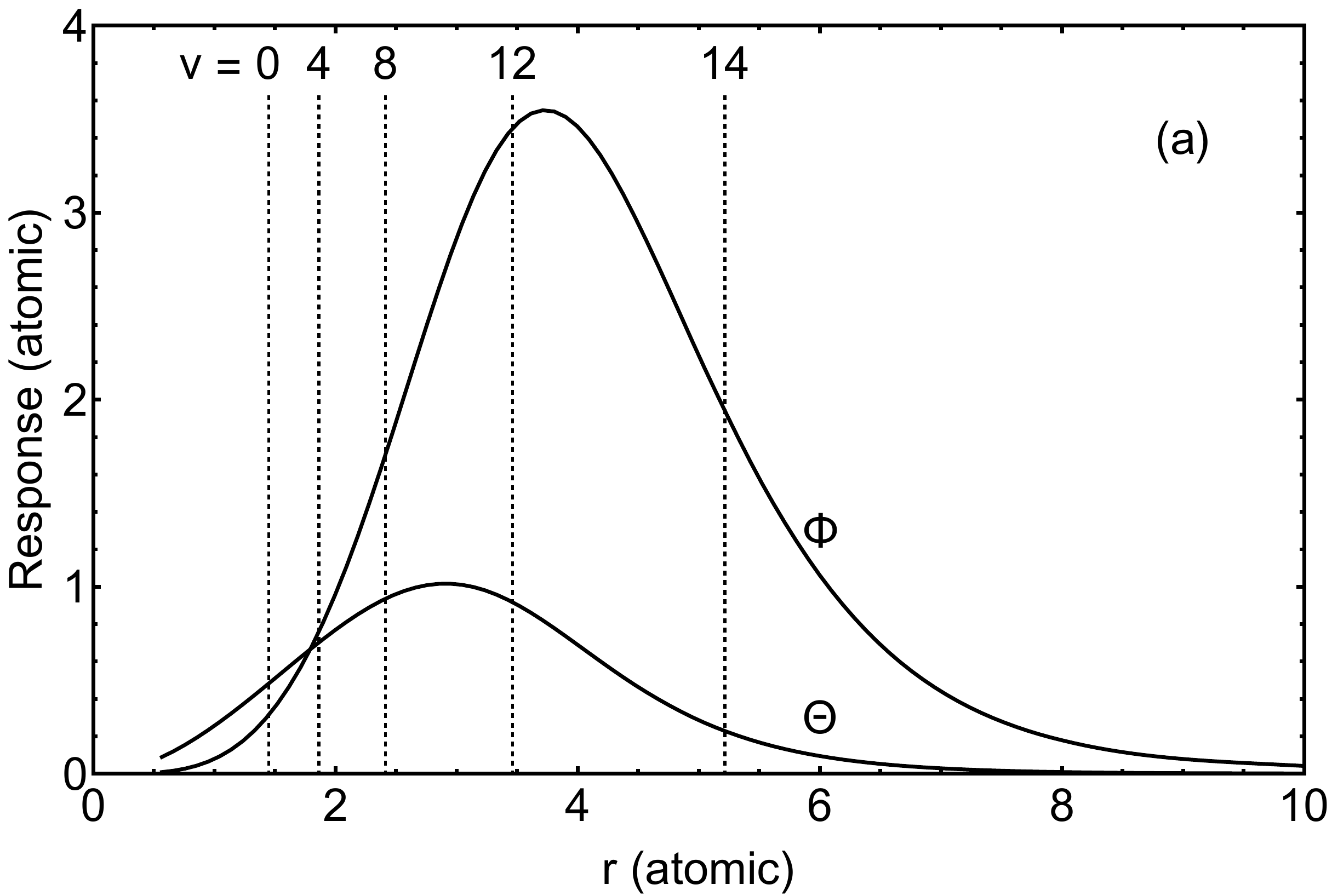}{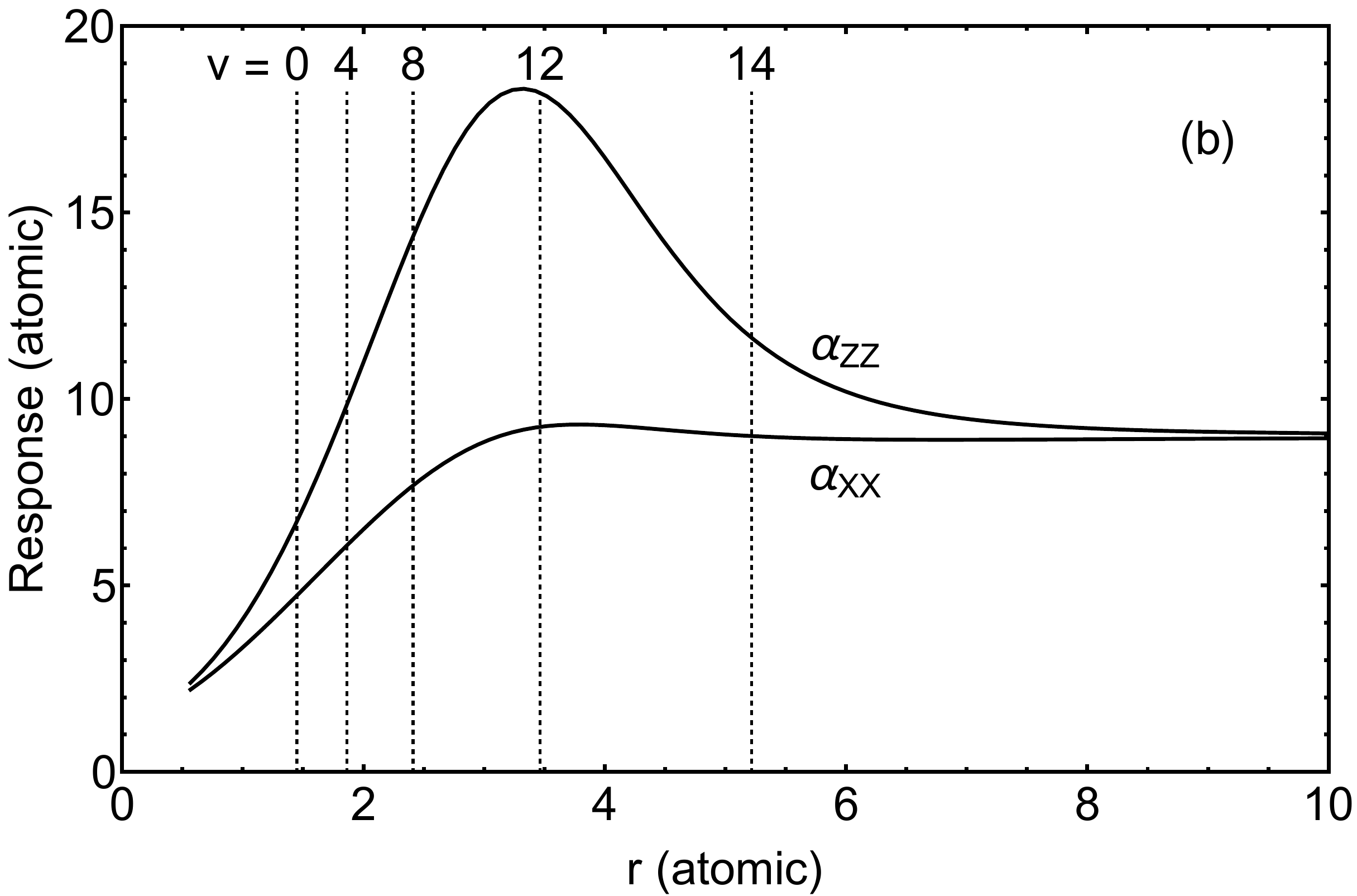}
\plottwo{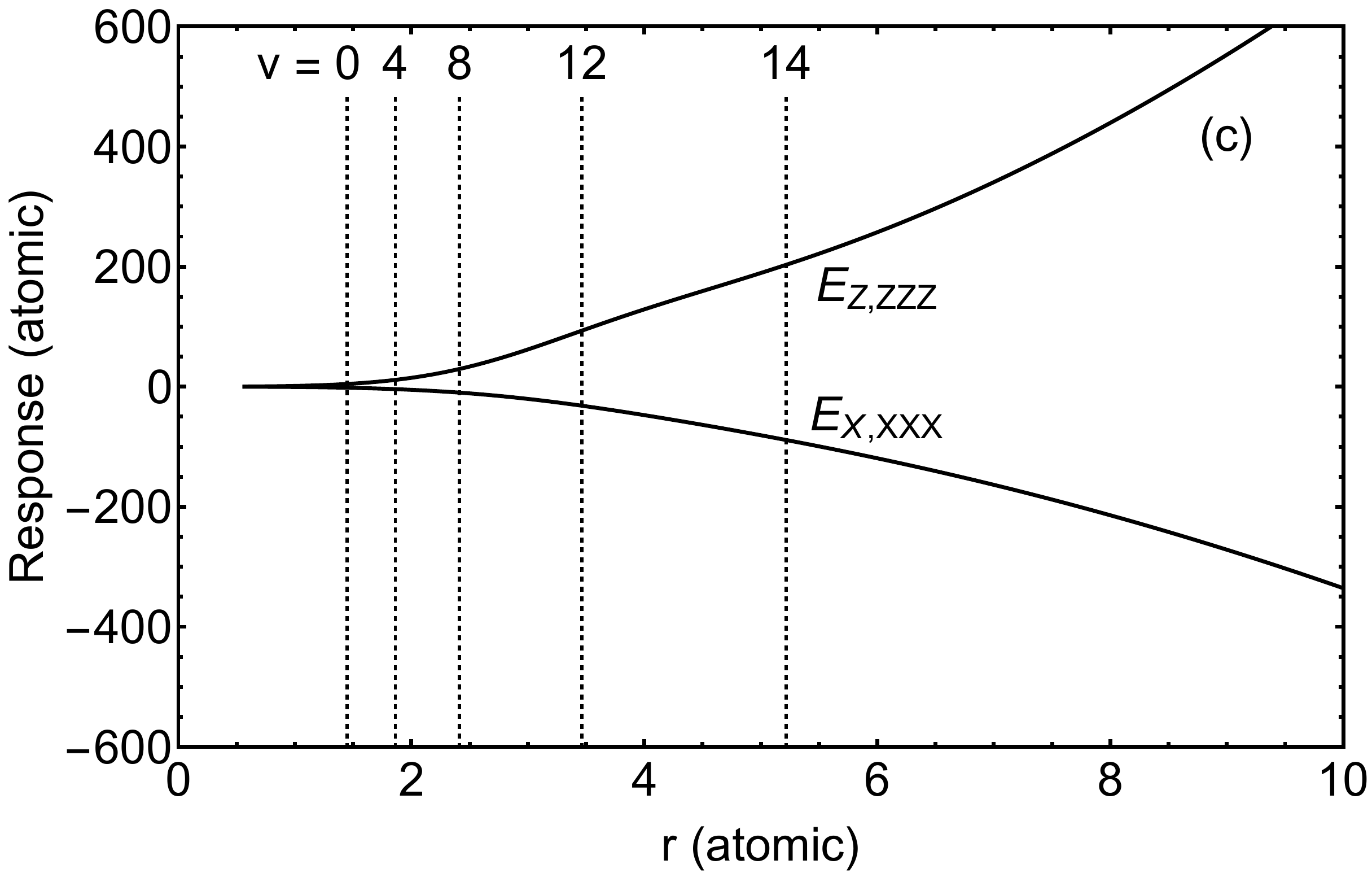}{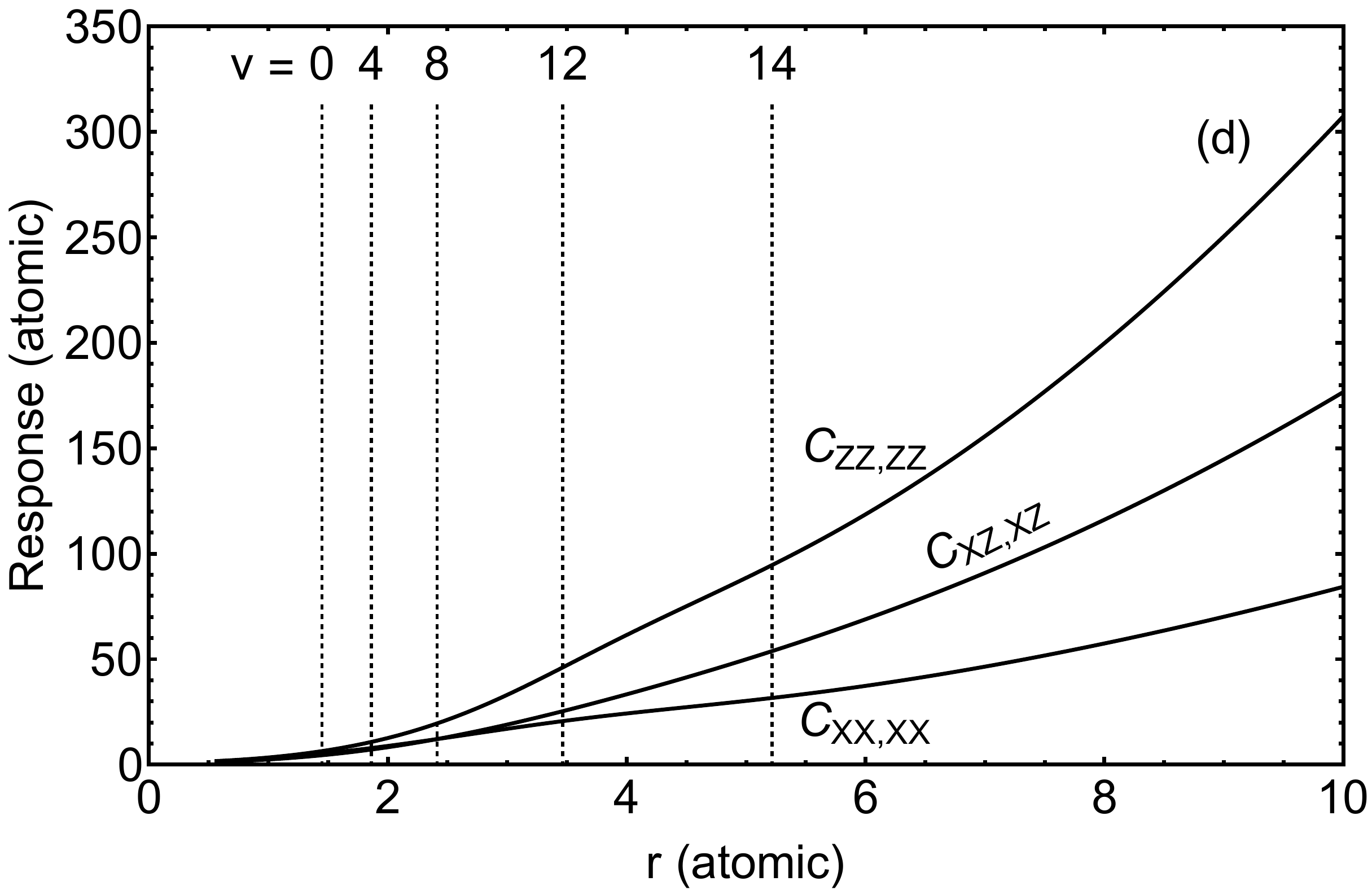}
\plottwo{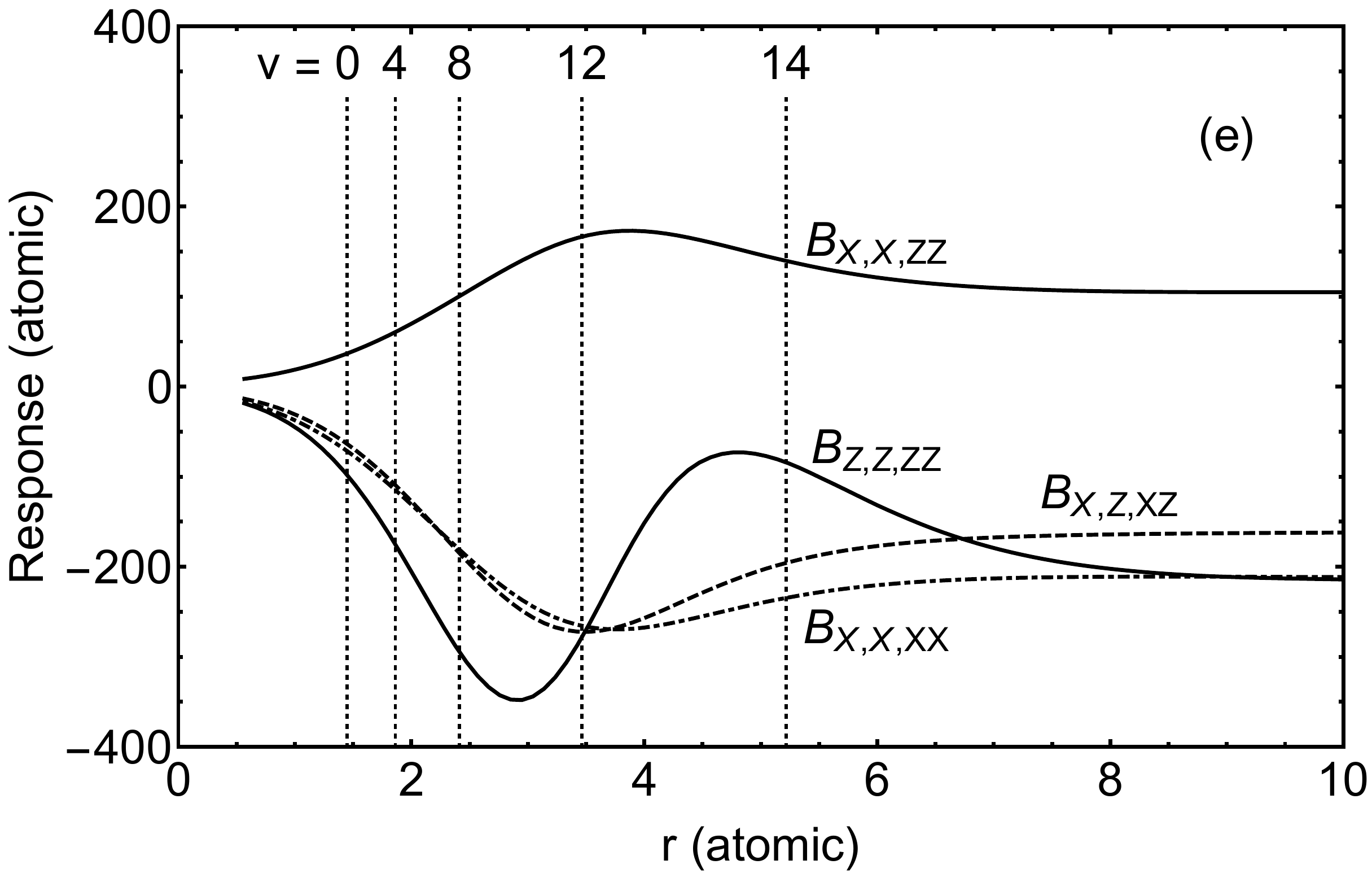}{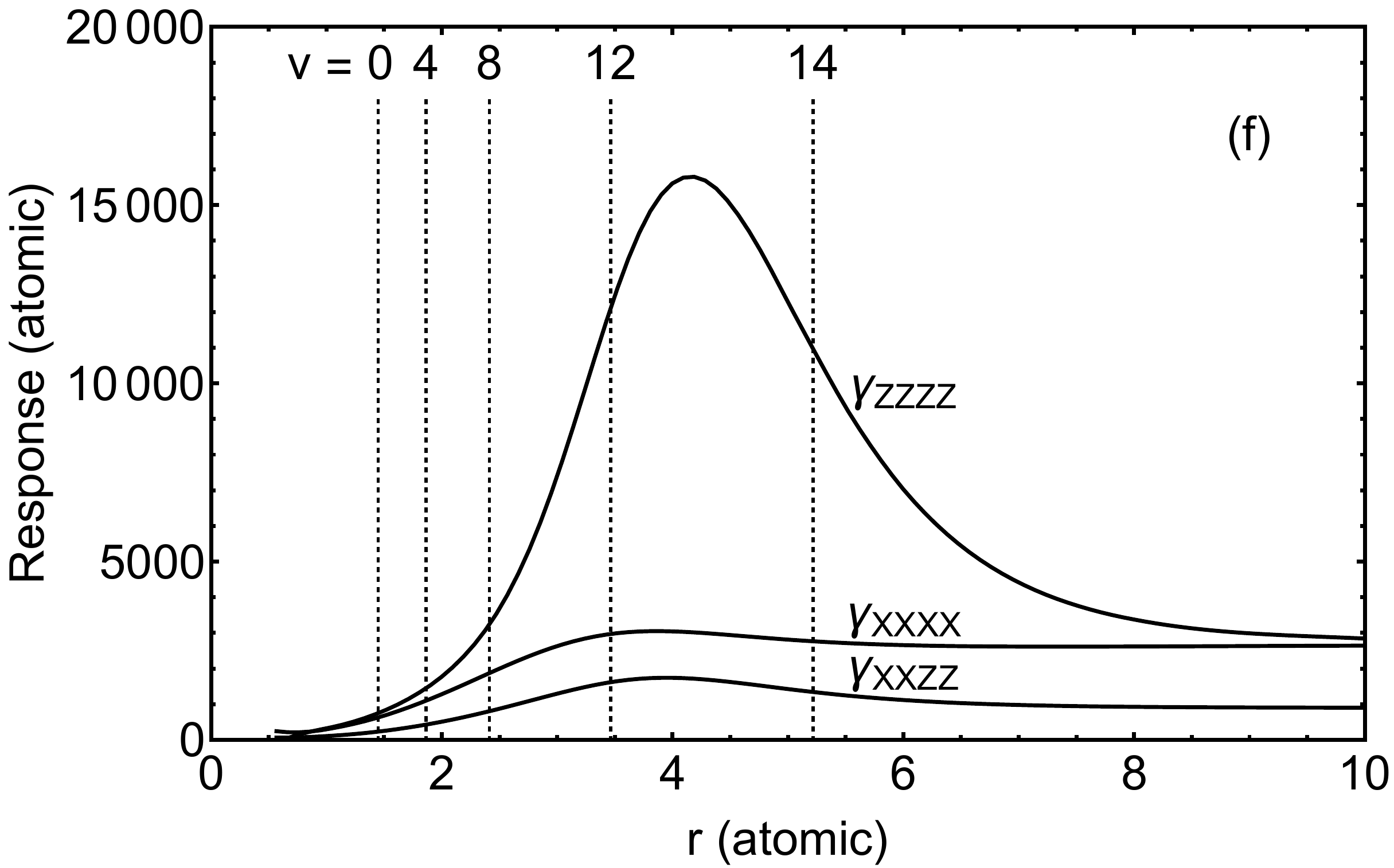}
\vskip-0.1truecm
\caption{The sixteen independent components of the electrical response tensors of \htwo, up to fourth rank, evaluated by \citet{miliordos2018} and used in this paper. The cartesian coordinate system $(X,Y,Z)$ used here is oriented such that the $Z$-axis is aligned with the internuclear separation vector. Panel (a): the quadrupole $\Theta\equiv\Theta_{_{\!Z\!Z}}$, and the hexadecapole $\Phi\equiv\Phi_{_{\!Z\!Z\!Z\!Z}}$. Panel (b): the dipole polarizability, $\alpha$. Panel (c): the dipole-octupole polarizability ${\rm E}$. Panel (d): the quadrupole-quadrupole polarizability ${\rm C}$. Panel (e): the dipole-dipole-quadrupole polarizability ${\rm B}$. Panel (f): the second dipole hyperpolarizability $\gamma$.  Vertical, dotted lines in each panel show the expected internuclear separations for states with $j=0$ and vibrational quantum numbers as labelled.}
\label{fig:tensorcomponents}
\medskip
\end{figure}
\section{Graphs of the electrical response tensors of \htwo}
In this Appendix we show the electrical response tensors evaluated by \citet{miliordos2018}: sixteen independent components, each plotted as a function of the internuclear separation.

\section{Hamiltonian matrix elements of the perturbation}
It is convenient to expand the angular structure of the perturbation in spherical harmonics, $Y_{l,n}$. For the electric field configurations considered in this paper, which are both axisymmetric, the only non-zero contributions are from the order $n=0$. Our description of the electrical response of the molecule extends up to the hexadecapole, so we need only consider degrees $0\le l\le4$, and the odd terms (dipole, octupole) are guaranteed to be zero because the \htwo\ molecule is symmetric under inversion through the centre-of-mass. The angular structure of the perturbation is therefore 
\be
\Delta E(r,\theta,\phi)=a_{0,0}(r)\,Y_{0,0}(\theta,\phi) + a_{2,0}(r)\,Y_{2,0}(\theta,\phi) + a_{4,0}(r)\,Y_{4,0}(\theta,\phi),
\ee
where $(\theta,\phi)$ are the polar angles of the internuclear axis of the molecule. The coefficients $a_{0,0}$, $a_{2,0}$ and $a_{4,0}$ are functions of the internuclear separation, and depend on the electric field and its first three derivatives -- terms up to the hexadecapole, in a multipole expansion of the potential -- but are independent of $(\theta,\phi)$. Integration over the sphere thus allows us to determine the coefficients from
\be
a_{l,n}=\int_{4\pi}{\rm d}\Omega\;[Y_{l,n}(\theta,\phi)]^*\;\Delta E,
\ee
because the spherical harmonics are orthonormal.

To compute the matrix elements of the Hamiltonian perturbation in the unperturbed Hamiltonian basis, \ket{v,j,m}, we first evaluate matrix elements of the spherical harmonics in the angular momentum basis: $\element{j,m}{l,n}{j^\prime,m^\prime}\equiv\element{j,m}{Y_{l,n}}{j^\prime,m^\prime}$. These are almost all zero. In general they can be calculated from
\be
\element{j,m}{l,n}{j^\prime,m^\prime}=
\left[ {{(2j+1)(2j^\prime+1)(2l+1)}\over{4\pi}} \right]^{1/2}
{1\over{(-1)^{m^\prime}}}
\left(\begin{array}{ccc}
j & j^\prime & l \\
0 & 0        & 0
\end{array}\right)
\left(\begin{array}{ccc}
j & j^\prime & l \\
m & -m^\prime & n
\end{array}\right),
\ee
where the last two factors are Wigner-3J coefficients (which are related to the Clebsch-Gordan coefficients). The non-zero elements lying on or above the leading diagonal are: \element{j,m}{0,0}{j,m}, for $j\ge0$; \element{j,m}{2,0}{j,m}, for $j\ge1$; \element{j,m}{2,0}{j+2,m}, for $j\ge0$; \element{j,m}{4,0}{j,m}, for $j\ge2$; \element{j,m}{4,0}{j+2,m}, for $j\ge1$; and, \element{j,m}{4,0}{j+4,m}, for $j\ge0$. Those below the leading diagonal can be determined by symmetry.

Although it is straightforward to obtain explicit expressions for these matrix elements in terms of $j$ and $m$ we do not present those forms here. Instead we recommend that readers use equation (B3), because functions for rapid evaluation of the Wigner-3J coefficients are readily available in various software packages and programming languages.

\subsection{Expansion coefficients for the case of a uniform field}
In the case of a uniform electric field, $F$, the only contributions to the electrical perturbation come from the dipole polarizability, $\alpha$, and the second dipole hyperpolarizability, $\gamma$. The coefficients in equation (B1) in this case are:
\be
a_{0,0}=-{1\over3}\sqrt{\pi}F^2\left\{\azz+2\axx + {1\over{60}}(3\gzzzz+12\gxxzz+8\gxxxx)F^2\right\},
\ee
and
\be
a_{2,0}=-{1\over 3}\sqrt{{{\pi}\over{5}}}F^2\left\{2\azz-2\axx
+ {1\over{21}}(3\gzzzz+3\gxxzz-4\gxxxx)F^2\right\},
\ee
and
\be
a_{4,0}=- {{2\sqrt{\pi}}\over{315}}(\gzzzz-6\gxxzz+\gxxxx)F^4.
\ee
To avoid unnecessary clutter in these expressions, we have not explicitly shown the dependence of the tensor components (\azz, etc.) on the internuclear separation, $r$. 

\subsection{Expansion coefficients for the case of a point-like elementary charge}
In the field of a point-like charge none of the derivatives of the electrical potential vanish, so there are contributions to the electrical perturbation associated with all seven response tensors. Let $R$ denote the separation between the charge and the centre-of-mass of the molecule, then the expansion coefficients in this case are:
\begin{eqnarray}
a_{0,0}=&-{1\over3}\sqrt{\pi}{{q^2}\over{R^4}}\Big\{\azz+2\axx
+{3\over 5}(\czzzz+8\cxzxz+8\cxxxx){1\over{R^2}}\hskip 4.5cm \nonumber\\
&-{2\over 5}(\bzzzz+\bxxzz+4\bxzxz+4\bxxxx){{q}\over{R^3}}
+{1\over{60}}(3\gzzzz+12\gxxzz+8\gxxxx){{q^2}\over{R^4}} \Big\},
\end{eqnarray}
and
\begin{eqnarray}
a_{2,0}=&\sqrt{{{\pi}\over{5}}}{{q}\over{R^3}}\Big\{ 2\Theta 
-{2\over 3}(\azz-\axx){{q}\over{R}}
-{2\over 7}\left( 3\ezzzz-8\exxxx
+5\czzzz+4\cxzxz-8\cxxxx\right){{q}\over{R^3}}\hskip0.5cm\nonumber\\
&+{1\over{21}}(11\bzzzz+2\bxxzz+8\bxzxz-16\bxxxx){{q^2}\over{R^4}}
-{1\over{63}}(3\gzzzz+3\gxxzz-4\gxxxx){{q^3}\over{R^5}}\Big\},\hskip0.5cm
\end{eqnarray}
and
\begin{eqnarray}
a_{4,0}=&{2\sqrt{\pi}}{{q}\over{R^5}}\Big\{ {1\over 3}\; \Phi
-{{4}\over{21}}(\ezzzz+2\exxxx){{q}\over{R}}
-{{4}\over{35}}(2\czzzz-4\cxzxz+\cxxxx){{q}\over{R}}\hskip2cm\nonumber\\
&+{{2}\over{105}}(3\bzzzz-2\bxxzz-8\bxzxz+2\bxxxx){{q^2}\over{R^2}}
-{{1}\over{315}}(\gzzzz-6\gxxzz+\gxxxx){{q^3}\over{R^3}}\Big\}. \hskip0.3cm
\end{eqnarray}

\section{Transition dipole matrix elements}
To determine the transition dipole matrix elements it is convenient to expand the electric dipole moment operator of the perturbed molecule (8) in spherical harmonics --- just as we did for the Hamiltonian of the perturbation in Appendix B. Rather than using the cartesian form $\mu_\alpha=(\mu_x,\mu_y,\mu_z)$ of the dipole moment, which follows directly from equation (\ref{eq:dipolemomentspecificexpression}), it is simplest to work with the spherical harmonic representation of that vector,
\be
\left\{\mu_{-1},\mu_0,\mu_{+1}\right\}=\left\{  {1\over{\sqrt{2}}}\left(\mu_x-i\,\mu_y\right),\mu_z,-{1\over{\sqrt{2}}}\left(\mu_x+i\,\mu_y\right) \right\}
\ee
and form the spherical harmonic expansion of each of those components. The expressions are given below for the two field configurations studied in this paper.

\subsection{Expansion coefficients for the case of a uniform field}
In the case of a uniform electric field, $F$, the only contributions to the electric dipole come from the dipole polarizability, $\alpha$, and the second dipole hyperpolarizability, $\gamma$. The spherical harmonic expansion coefficients in this case are:
\be
a_{0,0}(0)= {\sqrt{\pi}\over{3}}F\Big\{ 2\azz+4\axx +{1\over{15}}(3\gzzzz+12\gxxzz+8\gxxxx)F^2\Big\},
\ee
\be
a_{2,0}(0)= {4\over 3}\sqrt{{{\pi}\over{5}}}F\Big\{\azz-\axx+ {1\over{21}}(3\gzzzz+3\gxxzz-4\gxxxx)F^2 \Big\},
\ee
\be
a_{4,0}(0)={8\over{315}}\sqrt{\pi}\;F^3\,\Big\{\gzzzz-6\gxxzz+\gxxxx\Big\},
\ee
\be
a_{2,1}(1)=a_{2,-1}(-1)=\sqrt{{{\pi}\over{15}}}F \Big\{ 2\azz-2\axx + {1\over{21}}(3\gzzzz+3\gxxzz-4\gxxxx)F^2 \Big\},
\ee
and
\be
a_{4,1}(1)=a_{4,-1}(-1)= {2\over{63}} \sqrt{{{2\pi}\over{5}}}\; F^3\,\Big\{\gzzzz-6\gxxzz+\gxxxx\Big\},
\ee
where $a_{l,n}(m)$ denotes the $(l,n)$ coefficient of the spherical harmonic expansion of the $\mu_m$ component.

\subsection{Expansion coefficients for the field of a point-like elementary charge}
In the field of a point-like charge there are contributions to the dipole induced by the octupole of the applied field, and by the dipole-quadrupole of the applied field. The expansion coefficients in this case are:
\begin{eqnarray}
a_{0,0}(0)=&{1\over3} \sqrt{\pi} {{q}\over{R^2}} \Big\{ 2\azz+4\axx
-{4\over{5}}(\bzzzz+\bxxzz+4\bxzxz+4\bxxxx){{q}\over{R^3}}\nonumber \\
&+{1\over{15}}(3\gzzzz+12\gxxzz+8\gxxxx){{q^2}\over{R^4}} \Big\},\hskip 5.5cm
\end{eqnarray}
\begin{eqnarray}
a_{2,0}(0)=&2\sqrt{{{\pi}\over{5}}}{{q}\over{R^2}}\Big\{ {2\over 3}(\azz-\axx)
+{1\over 7}(3\ezzzz-8\exxxx){{1}\over{R^2}}\hskip7cm\\
&-{1\over{21}}(11\bzzzz+2\bxxzz+8\bxzxz-16\bxxxx){{q}\over{R^3}}
+{2\over{63}}(3\gzzzz+3\gxxzz-4\gxxxx){{q^2}\over{R^4}}\Big\},\hskip 0.3cm\nonumber
\end{eqnarray}
\begin{eqnarray}
a_{4,0}(0)=
&{{8}\over{21}}\sqrt{\pi}{{q}\over{R^4}}\Big\{\ezzzz+2\exxxx
-{{1}\over{5}}(3\bzzzz-2\bxxzz-8\bxzxz+2\bxxxx){{q}\over{R}}\hskip 1cm\nonumber\\
&+{{1}\over{15}}(\gzzzz-6\gxxzz+\gxxxx){{q^2}\over{R^2}}\Big\},\hskip8cm
\end{eqnarray}
\begin{eqnarray}
a_{2,1}(1)=&a_{2,-1}(-1)\hskip12.5cm\nonumber\\
=&\sqrt{{{\pi}\over{15}}}{{q}\over{R^2}}\Big\{ 2\azz-2\axx
-{2\over 7}(3\ezzzz-8\exxxx){{1}\over{R^2}}\hskip6cm\\
&-{2\over{7}}(\bzzzz-3\bxxzz+2\bxzxz-4\bxxxx){{q}\over{R^3}}
+{1\over{21}}(3\gzzzz+3\gxxzz-4\gxxxx){{q^2}\over{R^4}}\Big\},\nonumber
\end{eqnarray}
and
\begin{eqnarray}
a_{4,1}(1)=&\;a_{4,-1}(-1)\hskip13.8cm\nonumber\\
=&{2\over{21}}\sqrt{{{2\pi}\over{5}}}{{q}\over{R^4}}\Big\{
5\ezzzz+10\exxxx
-(3\bzzzz-2\bxxzz-8\bxzxz+2\bxxxx){{q}\over{R}}\hskip2cm\\
&+{1\over3}(\gzzzz-6\gxxzz+\gxxxx){{q^2}\over{R^2}}\Big\}.\hskip9.6cm\nonumber
\end{eqnarray}

\end{document}